\documentclass[lettersize,journal]{IEEEtran}
\usepackage{color}
\usepackage{amsmath,graphicx,amsfonts,multirow,booktabs,float,threeparttable,lipsum,dsfont,enumerate,caption}
\usepackage{blindtext}
\usepackage{diagbox}
\usepackage{tablefootnote}
\usepackage[table]{xcolor}
\usepackage{algorithm}
\usepackage{algpseudocode}
\usepackage{pifont}
\usepackage{amssymb}
\usepackage{tikz}
\usepackage{textpos,pgfplots,adjustbox,pifont}
\usepgfplotslibrary{groupplots}
\usepackage[hidelinks]{hyperref}
\usepackage{cite}

\definecolor{myBlue}{HTML}{4477AA}
\definecolor{myRed}{HTML}{BB5566}
\definecolor{myGreen}{HTML}{228833}
\definecolor{myYellow}{HTML}{DDAA33}

\newcommand{\tabincell}[2]{\begin{tabular}{@{}#1@{}}#2\end{tabular}}

\ifCLASSOPTIONcompsoc
    \usepackage[caption=false, font=normalsize, labelfont=sf, textfont=sf]{subfig}
\else
\usepackage[caption=false, font=footnotesize]{subfig}
\fi

\def\x{{\mathbf{x}}}
\def\z{{\mathbf{z}}}
\def\p{{\mathbf{p}}}
\def\r{{\mathbf{r}}}
\def\cc{{\mathbf{c}}}
\def\u{{\mathbf{u}}}
\def\vv{{\mathbf{v}}}

\begin{document}

\title{Self-supervised Reflective Learning through Self-distillation and Online Clustering\\for Speaker Representation Learning}

\author{Danwei~Cai, Zexin~Cai, Ze Li, 
        and~Ming~Li,~\IEEEmembership{Senior Member,~IEEE}%
\thanks{Danwei Cai and Zexin Cai are with the Department of Electrical and Computer Engineering, Duke University, Durham, NC, 27705, USA, e-mail: \{danwei.cai, zexin.cai\}@duke.edu}%
\thanks{Ze Li and Ming Li are with Suzhou Municipal Key Laboratory of Multimodal Intelligent Systems, Digital Innovation Research Center, Duke Kunshan University, Kunshan, China and School of Computer Science, Wuhan University, Wuhan China,
e-mail: \{ze.li, ming.li369\}@dukekunshan.edu.cn}
\thanks{Corresponding author: Ming Li.}%
}


\maketitle

\begin{abstract}
  Speaker representation learning is crucial for voice recognition systems, with recent advances in self-supervised approaches reducing dependency on labeled data.
  Current two-stage iterative frameworks, while effective, suffer from significant computational overhead due to repeated rounds of clustering and training.
  They also struggle with noisy pseudo labels that can impair model learning.
  This paper introduces self-supervised reflective learning (SSRL), an improved framework that addresses these limitations by enabling continuous refinement of pseudo labels during training.
  Through a teacher-student architecture and online clustering mechanism, SSRL eliminates the need for iterative training rounds.
  To handle label noise, we incorporate noisy label modeling and pseudo label queues that maintain temporal consistency. Experiments on VoxCeleb show SSRL's superiority over current two-stage iterative approaches, surpassing the performance of a 5-round method in just a single training round.
  Ablation studies validate the contributions of key components like noisy label modeling and pseudo label queues.
  Moreover, consistent improvements in pseudo labeling and the convergence of cluster counts demonstrate SSRL's effectiveness in deciphering unlabeled data. This work marks an important advancement in efficient and accurate self-supervised speaker representation learning through the novel reflective learning paradigm.  
\end{abstract}


\section{Introduction}
\IEEEPARstart{S}{peaker} representation learning is a core component of voice recognition systems that aims to extract discriminative speaker characteristics from speech signals.
Recent research has demonstrated significant advances in self-supervised learning approaches for speaker representation learning \cite{chen_large-scale_2022, chen_pushing_2023, tu2024contrastive, liu2023self, 10448455,fathan2024analytic,wang2024leveraging}.
These methods enable the utilization of unlabeled data, reducing dependency on manual annotation and facilitating deployment in practical applications.

Our previous work introduced a two-stage iterative framework for unsupervised speaker representation learning \cite{cai_iterative_2021, 9741340}.
The first stage performs self-supervised speaker representation learning, while the second stage combines clustering with discriminative training.
In this framework, clustering algorithms analyze the learned representations to generate pseudo labels for unlabeled data.
These pseudo labels, despite containing some noise, are then used to train the network through discriminative learning.
The process iterates, with each training round refining the representations and improving the quality of pseudo labels, leveraging DNNs' inherent robustness to label noise.

However, this two-stage iterative approach, while effective, introduces significant computational overhead.
The primary limitation stems from the iterative process of generating pseudo labels through clustering followed by discriminative training.
Furthermore, the initial pseudo labels derived from clustering contain substantial noise, which impairs the model's ability to learn discriminative speaker features.

\begin{figure}
    \centering
    \includegraphics[width=0.55\linewidth]{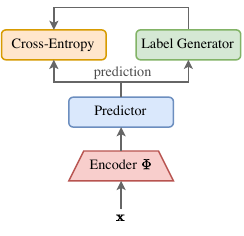}
    \caption{A naive solution to bypass the iterative process of clustering and discriminative training in two-stage framework.}
    \label{fig: naive solution}
\end{figure}

To streamline learning and enhance efficiency, we sought to bypass the iterative nature of the two-stage framework by updating pseudo labels at every training step rather than after each full training round.
As shown in Figure \ref{fig: naive solution}, A naive solution is to directly generate labels from model predictions -- e.g., assigning the class with the highest posterior probability.
However, this approach can lead to a degenerate solution, where the model suffers from confirmation bias, collapsing to a single prediction across all samples \cite{chen_exploring_2021}.

The key to avoiding this degenerate solution is to decouple training from label assignment, ensuring the model does not directly train on its own predictions.
In the two-stage iterative framework, this decoupling is achieved by clustering pseudo labels after model convergence.
To generate pseudo label at every training step while preventing degenerate solution, we propose self-supervised reflective learning (SSRL), which achieves this decoupling via self-supervised knowledge distillation \cite{hinton_distilling_2015, grill_bootstrap_2020, chen_exploring_2021, caron_emerging_2021}.
The framework employs a teacher model to generate pseudo labels for training a student model.
The teacher model, updated as an exponential moving average (EMA) of the student, functions as an ensemble of past model states, thereby stabilizing pseudo label generation and preventing overfitting.
This mechanism enables continuous knowledge integration from previous training steps while maintaining stable learning dynamics.
Essentially, the student model learns from its own reflection -- a feedback-driven process where insights from previous training steps guide and improve future learning.

Another limitation of the two-stage iterative framework is the high noise in pseudo labels, which degrades learning. We address this with two key strategies.
First, we maintain a queue of historical pseudo labels to filter out outlier predictions and ensures consistent and reliable pseudo labels.
Second, we integrate a label noise modeling strategy using a two-component Gaussian mixture model (GMM) to capture the loss distribution of training samples, as described by \cite{arazo_unsupervised_2019}. This approach leverages the tendency of DNNs to prioritize correct targets, providing a clean label probability for each sample, which is then used to adjust the final loss.

This paper presents several contributions to the field of self-supervised speaker representation learning:
\begin{enumerate}
\item The introduction of a new learning paradigm, `self-supervised reflective learning', that bridges self-supervised representation learning with self-supervised knowledge distillation and online clustering, eliminating iterative bottlenecks of the two-stage unsupervised framework.
\item A detailed examination of how noisy label modeling, when combined with self-supervised knowledge distillation, can handle label noise, thereby improving the robustness of the learning process.
\item Our approach surpasses existing iterative methodologies, marking a significant step forward in efficient and accurate unsupervised speaker representation learning.
\end{enumerate}

\section{Related works}

\subsection{Self-supervised knowledge distillation}
In self-supervised learning, knowledge distillation involves processing two distinct views through separate encoders and mapping one to the other using a predictor.
A potential pitfall is the convergence of outputs to a uniform constant.

To addressed this issue, `bootstrap your own latent' (BYOL) introducing a momentum-based teacher network to generate targets for the student network \cite{grill_bootstrap_2020}.
Both networks process distinct views of the same instance through data augmentation.
The student network aligns its outputs with the teacher network using a predictor, while the teacher network is updated through an EMA of the student network's weights.
He \textit{et al.} later introduced the `simple siamese' (SimSam) approach, which simplified BYOL by removing the momentum mechanism \cite{chen_exploring_2021}, showing that while EMA wasn't essential, it could enhance performance.

Similar to BYOL, `self-distillation with no labels' (DINO) focuses on regression from student to momentum encoder representations \cite{caron_emerging_2021}.
However, DINO differs by using cross-entropy instead of Mean square error (MSE) or cosine similarity for alignment, and by centralizing the teacher's output using a running mean with temperature-scaled softmax.
With its large output dimension (65536 in the original paper), DINO effectively functions as an online clustering mechanism.

Our proposed SSRL approch deviates from the DINO approach in multiple ways.
Firstly, SSRL builds on a two-stage iterative unsupervised framework, starting with more reliable pseudo labels rather than random initialization.
Unlike DINO's continuous class probability distributions, SSRL delivers discrete class predictions thus enables discriminative training of the model.
Additionally, SSRL adopts noisy student training where only the student network undergoes augmentation and noise, allowing the teacher to generate higher quality pseudo labels.
Lastly, SSRL integrates a pseudo label queue and noisy label modeling handle label inaccuracies.

Self-supervised knowledge distillation has also been applied to speech representation learning \cite{elbanna2022byol, liu2024dinosr, 10095373}. For instance, DinoSR \cite{liu2024dinosr} adapts DINO for speech representation learning with sequence modeling, combining masked language modeling (MLM), self-distillation, and online clustering. While DinoSR focuses on frame-level pretraining using an explicit codebook for pseudo-label generation, our work targets utterance-level modeling of speaker characteristics.

\subsection{Self-supervised pseudo labeling}
In self-supervised learning, many approaches use clustering-derived pseudo labels for discriminative training.
A primary approach, known as deep clustering (DC), combines conventional clustering with classification loss for network training \cite{caron2018deep}.
However, DC faces several challenges:
Firstly, the conventional off-the-shelf clustering requires feature extraction across the full dataset for every epoch.
Secondly, the clustering alters cluster indexes across epochs, necessitating a reset of the parametric classifier, leading to unstable network training.
Thirdly, The combination of discriminative and clustering losses can lead to degenerate solutions where all samples map to the same pseudo label \cite{asano_self-labelling_2020}.
DC avoids the issue by optimizing only one loss and keeping the other loss fixed between training epochs.

Prototypical contrastive learning (PCL) addresses the cluster index permutation by replacing the classification layer with cluster centroids \cite{li_prototypical_2021}.
It generates class probabilities by contrasting samples with centroids and incorporates instance discrimination.
However, PCL still requires per-epoch feature extraction and faces challenges with the divergence between evolving representations and fixed centroids.

Online deep clustering (ODC) improves upon DC using sample and centroid memories for pseudo label generation \cite{zhan_online_2020}.
It updates sample memory through moving averages and assigns labels based on nearest centroids.
To prevent degenerate solutions, ODC implements 'merge-and-split' operations and loss re-weighting for small clusters.

Recently, Asano \textit{et al.} introduced a self-labeling algorithm (SeLa) addressing the degenerate solutions in combined clustering and representation learning \cite{asano_self-labelling_2020}.
Contrary to DC's direct clustering application, SeLa determines label assignments $q(y|\x_i)$ from the network-derived class posterior probabilities $p(y|\x_i)$.
Here, $y$ denotes labels and $\x_i$ denotes data samples.
SeLa solves the cross-entropy optimization problem $\min_{q}\sum_i\sum_yq(y|\x_i)\log p(y|\x_i)$ with equal-sized partition constraints using the Sinkhorn-Knopp algorithm.
While this unifies network training and clustering objectives, SeLa's offline cluster assignment limits its scalability.

SwAV (Swapping Assignments between Views) advanced SeLa by introducing online clustering through optimal transport within mini-batches \cite{caron_unsupervised_2020}. Instead of processing the full dataset, SwAV enforces consistency between cluster assignments from multiple augmented views of the same image. The swapped prediction task -- predicting one view's code from another's representation -- enables online training while maintaining equipartition constraints via the Sinkhorn-Knopp algorithm. Chang \textit{et al.} apply SwAV to self-supervised speech representation learning by processing original speech and speaker-perturbed versions through shared encoders to create two views \cite{chang23_interspeech}.

Unlike the aforementioned methods which often rely on pseudo labels generated from randomly initialized feature representations, our approach harnesses the strength of a pre-trained self-supervised model.
This foundational difference enables our clustering module to produce semantic pseudo labels, sidestepping the pitfalls of arbitrary cluster assignments.

\subsection{Self-training in semi-supervised learning and unsupervised domain adaptation}

In semi-supervised learning scenarios, where there exists a limited labeled dataset complemented by a larger pool of unlabeled data, the objective is to harness the intrinsic structures or patterns prevailing within the unlabeled data to enhance the learning algorithm \cite{zhu2022introduction}.
A prevalent approach to realizing this objective is self-training \cite{9941371, amini2022self, ke2019dual, sohn2020fixmatch, chen2021semi, xie_self-training_2020}.
This method operates iteratively: initially, a model is trained using the available labeled data, serving as a basis to generate predictions on the unlabeled dataset.
Predictions made with high confidence, termed pseudo labels, are integrated into the training set, forming an augmented dataset on which the model undergoes further training.
This iterative cycle continues until convergence is achieved or a set number of iterations are completed.
The core intent of self-training is to exploit the inherent but latent structures within the unlabeled data, thereby augmenting the model's capacity to generalize effectively.

In the context of unsupervised domain adaptation (UDA), the labeled data are derived from a specific source domain, whereas the unlabeled data are from a distinct, but related, target domain.
The pivotal challenge lies in adeptly fine-tuning the model, which has been preliminarily trained on the source domain, ensuring its optimized performance when applied to the target domain by effectively utilizing the unlabeled target data.
The self-training method, due to its intrinsic reliance on unlabeled data, finds substantial applicability in UDA, seamlessly aligning with its fundamental principles \cite{busto2018open, french2018self, zou2019confidence, zou2018unsupervised}.

Numerous variants of the self-training technique have been innovated for semi-supervised learning and unsupervised domain adaptation.
For instance, one variant employs the teacher-student architecture to impose a consistency regularization \cite{tarvainen2017mean, laine2016temporal}.
Here, metrics such as the MSE or Kullback-Leibler divergence are commonly used to apply prior constraint assumptions on the unlabeled data.
Central to consistency regularization is that the model's output remains robust to specific perturbations.
Moreover, there exist models like deep co-training \cite{qiao2018deep} and Tri-Net \cite{dong2018tri}, based on the disagreement-based paradigm.
These models foster the simultaneous training of multiple models, leveraging the disagreements among them as a critical aspect of the learning process.

In contrast to self-training, our proposed SSRL approach operates within a purely unsupervised setting, devoid of any reliance on labeled data.
Unlike self-training, where the set of labels is predetermined, there is no prior knowledge of class counts or explicit label information in the unsupervised setting.
The proposed SSRL method uses the teacher-student framework, and the teacher provides pseudo labels based on an online clustering mechanism.
This dynamic mechanism fosters the creation of evolving clusters, which are adaptable and capable of undergoing refinements throughout the learning process.
Thus, the SSRL method, with its capacity for dynamic clustering, promises enhanced performance and robustness in unsupervised learning scenarios.

\begin{figure*}[t]
     \centering
     \subfloat[Two-stage framework with iterative training.]{
    	\includegraphics{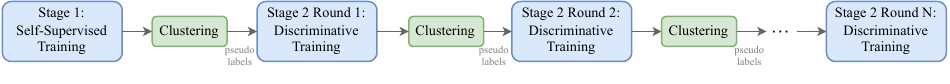}}
        \\
     \subfloat[Improved two-stage framework with SSRL.]{
    	\includegraphics{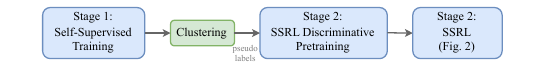}}
     \caption{A csomparison of the proposed SSRL method with our previously proposed two-stage method with iterative training.}
     \label{fig: comp}
\end{figure*}

\begin{figure}[t]
\centering
  \includegraphics[width=0.88\linewidth]{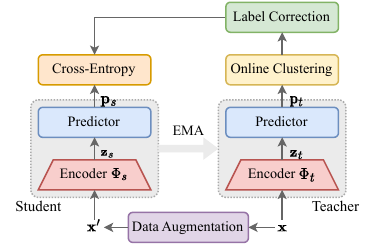}
  \caption{The proposed self-supervised reflective learning (SSRL) method.}
  \label{fig:selflab}
\end{figure}

\subsection{Speaker representation learning}
Before the emergence of deep learning, speaker representation learning primarily relied on statistical modeling techniques. Among these, the Gaussian Mixture Model - Universal Background Model (GMM-UBM) was widely used to model speaker features probabilistically \cite{reynolds2000speaker, gauvain1994maximum, kenny2007joint}. This approach was later enhanced by i-vector representations, which leveraged factor analysis to map speaker characteristics into a low-dimensional space, improving speaker recognition and verification performance \cite{dehak_front-end_2011}.

With the advent of deep neural networks, speaker modeling transitioned to data-driven feature extraction, enabling more robust and discriminative speaker embeddings. One of the earliest breakthroughs was the x-vector framework, which introduced Time-Delay Neural Networks (TDNNs) along with a statistics pooling layer to aggregate features at the utterance level \cite{7404779, snyder_x-vectors:_2018}. This significantly improved robustness to variable-length speech segments. Following this, ResNet-based architectures were applied to speaker modeling, incorporating convolutional layers to effectively capture local speaker features \cite{cai2018exploring, zhou_resnext_2021}. Further advancements led to ECAPA-TDNN, which introduced channel-wise attention mechanisms and multi-scale feature learning, enhancing speaker discriminability while maintaining a compact model size \cite{desplanques_ecapa-tdnn_2020}. More recently, Transformer-based architectures, such as Conformer, have been explored, integrating self-attention mechanisms with convolutional layers to better capture both global and local speaker characteristics \cite{zhang_mfa-conformer_2022, 10095433, 10572375}.

A major paradigm shift in speaker modeling has been the adoption of large-scale self-supervised pre-trained models. Models such as wav2vec 2.0 \cite{baevski_wav2vec_2020}, HuBERT \cite{hsu_hubert_2021}, and WavLM \cite{9814838} have been utilized as feature extractors and fine-tuned for speaker-related tasks \cite{fan_exploring_2021, vaessen_fine-tuning_2022, chen_large-scale_2022}. These approaches leverage self-supervised learning as a pretraining method, allowing for scalable and data-efficient speaker modeling. By learning from vast amounts of unlabeled speech data, these models significantly enhance generalizability and reduce reliance on manually labeled datasets.

\subsection{Two-stage iterative framework for unsupervised speaker representation learning}
Multi-stage unsupervised learning frameworks have been widely adopted in various domains, including hyperspectral image processing \cite{9714360, 10233913, 10504844}, where structured priors and iterative refinements enhance representation quality. In speaker representation learning, a two-stage iterative framework has been proposed to leverage large-scale unlabeled data efficiently \cite{cai_iterative_2021, 9741340, thienpondtidlab}. In the first stage, self-supervised methods are used to extract initial speaker embeddings and pseudo labels. The second stage involves an iterative discriminative training process to refine these embeddings.

To enhance stage one, more advanced self-supervised representation learning such as DINO are proposed \cite{10314722}. 
Zhao et al. \cite{zhao2024prototype} further refined DINO with the Prototype Division method, effectively mitigating speaker confusion and enhancing overall performance.
Addtionally, Tao et al. \cite{tao2023self} proposed a multi-modal contrastive learning technique using diverse positive pairs by cross-referencing speech and face data, improving the robustness of speaker encoders. 

A significant challenge in this framework is the presence of noisy pseudo labels. To address this, several methods have been developed.  Tao et al. \cite{tao_self-supervised_2022} introduced a technique that extracts reliable labels based on the neural network’s fitting ability during training. Han et al. \cite{han_self-supervised_2022} and Zhou et al. \cite{10448455} proposed using a Gaussian Mixture Model (GMM) to dynamically model loss distribution, distinguishing between reliable and unreliable labels, and correcting the unreliable ones using model predictions. Chen et al. \cite{chen_self-supervised_2023} developed a method that coordinates information between audio and visual modalities through an ``update by disagreement'' strategy, improving pseudo label quality by leveraging inter-modal disagreements.

While these methods improve pseudo label quality, they typically require iterative training stages. 
Fang et al. \cite{fang2024improving} employed a label ensemble approach to smoothly correct noisy speaker labels by the exponential moving average of model predictions at each training epoch. Similarly, the approach presented in this paper dynamically improves pseudo labels at each epoch, eliminating the need for multiple training rounds and enhancing both efficiency and accuracy in speaker representation learning.

\section{Methods}

This section introduces the self-supervised reflective learning (SSRL) approach, which improves the two-stage iterative framework \cite{cai_iterative_2021}. 
In the original two-stage framework, the first stage applies self-supervised representation learning and generates initial pseudo labels. 
The second stage consists of multiple training rounds, each comprising: (1) pseudo-label generation through clustering, and (2) discriminative training using these labels.

While maintaining the first stage unchanged, the proposed SSRL method replaces the multi-round process with continuous label refinement during a single training phase. 
As illustrated in Figure \ref{fig: comp}, SSRL requires an initialization step to ensure stable pseudo-label generation. 
This can be achieved either through brief discriminative training using Stage 1 pseudo labels, or by directly employing the Stage 1 self-supervised model as the encoder with a predictor initialized using pseudo cluster centroids. 
Following initialization, SSRL dynamically updates pseudo labels during training through its reflective learning mechanism.

Figure \ref{fig:selflab} illustrates the proposed SSRL method, with the detailed procedure outlined in Algorithm \ref{algo: ssrl}.

\begin{figure*}[ht]
\centering
\begin{minipage}{0.75\textwidth}
\begin{algorithm}[H]
\caption{Self-Supervised Reflective Learning (SSRL)}
\begin{algorithmic}[1]
\Require Unlabeled dataset $\mathcal{D}=\{\x_i | i=1,\cdots,N\}$; Initial pseudo labels $\mathcal{Y}=\{y_i| i=1,\cdots,N\}$; Teacher encoder $\Phi_t$ with parameters $\phi_t$; Teacher predictor $h_t$ with parameters $\psi_t$; Student encoder $\Phi_s$ with parameters $\phi_s$; Student predictor $h_s$ with parameters $\psi_s$
\Procedure{ReflectiveLearning}{$\mathcal{D}, \mathcal{Y}$}
    \State Train student network ($\Phi_s$ and $h_s$) with dataset $\{\mathcal{D}$, $\mathcal{Y}\}$ for $E_1$ epochs 
    \State Initialize $\Phi_t$ with $\Phi_s$ and $h_t$ with $h_s$ 
    \State $Q_i \leftarrow \text{Queue}(\text{length}=L)$ \Comment{Initialize empty pseudo label queue for all samples} 
    \State $p_\text{clean}(\ell_{t,i}) \leftarrow 1$ \Comment{Initialize clean label probability to 1 for all samples}
    \\
    \For{epoch in 1 to $E_2$}
        \For{batch $\mathcal{B}=\{\x_i|i=1,\cdots,B\}$ in $\mathcal{D}$}
        	\State Crop a \textbf{short} segment for each training sample $\mathcal{B}_s=\{\x_i'\}$
        	\State Apply data augmentation to $\mathcal{B}_s$
        	\State $\p_{s,i} \leftarrow \mathrm{softmax}(h_s\circ\Phi_s(\x_i'))$ \Comment{Student output}
        	\State $\mathcal{L} \leftarrow -\frac{1}{B}\sum_{i=1}^{B}p_{\text{clean}}(\ell_{t,i})\log p_s(y_i|\x_i')$ \Comment{Training loss}
        	\\
        	\State Crop a \textbf{long} segment for each training sample $\mathcal{B}_t=\{\tilde{\x}_i\}$
            \State $\p_{t,i} \leftarrow \mathrm{softmax}(h_t\circ\Phi_t(\tilde{\x}_i))$ \Comment{Teacher prediction}
            \State $y_i \leftarrow \mathrm{clustering}(\p_{t,i})$ \Comment{Online clustering}
            \State Enqueue $y_i$ to $Q_i$
            \State $y_i \leftarrow $ mode of labels in $Q_i$ \Comment{Label correction}
            \State $\ell_{t,i} \leftarrow -\log p_t(y_i|\tilde{\x}_i)$ \Comment{Cross entropy loss of teacher}
            \\
            \State Update student parameters using gradients from $\mathcal{L}$
			\State $\phi_t \leftarrow \lambda\phi_t+(1-\lambda)\phi_s$ \Comment{EMA update of teacher encoder}
			\State $\psi_t \leftarrow \lambda\psi_t+(1-\lambda)\psi_s$ \Comment{EMA update of teacher predictor}
        \EndFor
        \\
        \State Fit $\{\log \ell_{t,i}|i=1,\cdots,N\}$ with a GMM \Comment{Noisy label modeling}
        \State Update $p_\text{clean}(\ell_{t,i})$ using GMM
    \EndFor
\EndProcedure
\end{algorithmic}
\label{algo: ssrl}
\end{algorithm}
\end{minipage}
\end{figure*}

\subsection{Self-supervised knowledge distillation}
At the heart of our approach is the self-supervised knowledge distillation technique.
Given an unlabeled dataset, the teacher network generates cluster assignments which guide the training of the student network.
The teacher encoder, represented as $\Phi_t(\cdot)$, transforms the data sample $\x$ into a $D$-dimensional feature representation $\z_t \in \mathbb{R}^D$:
\begin{equation}
\z_t=\Phi_t(\x)
\end{equation}
Subsequently, a linear predictor, $h_t(\cdot)$, is employed to compute the probability distribution over $K$ clusters via a softmax operator. Let $p_t(k|\x)$ denotes the posterior probability that the sample $\x$ belongs to the $k^{\mathrm{th}}$ cluster, the vector $\p_t$ aggregates these probabilities for all $K$ clusters:
\begin{equation}
\p_t=\mathrm{softmax}(h_t(\z_t))=\mathrm{softmax}(h_t\circ\Phi_t(\x))
\end{equation}
where $p_t(k|\x)$ is the $k^{\mathrm{th}}$ element of $\p_t$. An online clustering mechanism then extracts cluster assignments $y \in \{1,2,\cdots,K\}$ from $\p_t$ for the training sample $\x$.

Following a parallel structure, the student encoder $\Phi_s(\cdot)$, coupled with the student predictor $h_s(\cdot)$ -- analogous in architecture to the teacher - produce the feature $\z_s$ and the class prediction $\p_s$ from another view of the same input $\x'$.
The student model's training utilizes the cross-entropy loss, under the supervision of the pseudo label $y$ derived from the teacher:
\begin{equation}
\mathcal{L} = -\frac{1}{N}\sum_{i=1}^{N}\log p_s(y_i|\x_i')
\end{equation}
where $N$ represents the number of data samples in a training batch.

\subsubsection{Enhancing the student's model capacity}
Drawing inspiration from the noisy student method in semi-supervised learning \cite{xie_self-training_2020}, our approach amplifies the student's modeling capacity by imposing noise into the training samples during the student's training.
Specifically, a short segment is extracted from the training utterance, followed by data augmentation techniques introducing background noise or convolutional reverberation to this segment.
Consequently, the student model processes these augmented snippets.
The teacher model, on the other hand, processes a longer clip of the same utterance in its unaltered form, facilitating the generation of stable pseudo labels.
Morever, other deep neural network training strategies can further improve the student's model capacity.
For instance, employing dropout can mitigate the risk of overfitting to the imprecise pseudo labels \cite{srivastava_dropout:_2014}.
Another approach involves the use of angular margin-based cross entropy \cite{deng_arcface_2019} as a loss function, fostering the student model to capture a more discerning feature space.

\subsubsection{Teacher model update mechanism}
Traditional knowledge distillation typically employs a teacher network, trained with labeled data and possessing superior model capacity.
However, under self-supervised settings, acquiring such a pre-trained teacher model is not feasible.
We hypothesize that the student model's capacity undergoes enhancement after each training cycle, courtesy of noisy student training.
As such, an advanced teacher model can be obtained by ensembling student models from previous training steps.
In specific terms, we employ an EMA technique on the student's parameters to refine the teacher model \cite{moco,grill_bootstrap_2020,caron_emerging_2021} .
Denoting the parameters of student encoder $\Phi_s(\cdot)$ as $\phi_s$ and the parameters of student predictor $h_s(\cdot)$ as $\psi_s$, the teacher's parameters $\phi_t$ and $\psi_t$ undergo an update as:
\begin{equation}\begin{aligned}
\label{eq: ema}
	\phi_t & \leftarrow \lambda\phi_t+(1-\lambda)\phi_s \\
	\psi_t & \leftarrow \lambda\psi_t+(1-\lambda)\psi_s
\end{aligned}\end{equation}
where $\lambda\in[0,1)$ serves as a momentum coefficient.
Through the EMA update mechanism, the teacher consistently outperforms the student during the training process, thereby facilitating the student's learning by providing pseudo labels of higher quality.

\begin{figure*}[th!]
  \centering
  \subfloat[Initial Model, Linear Scale]{
   \includegraphics[width=0.245\linewidth]{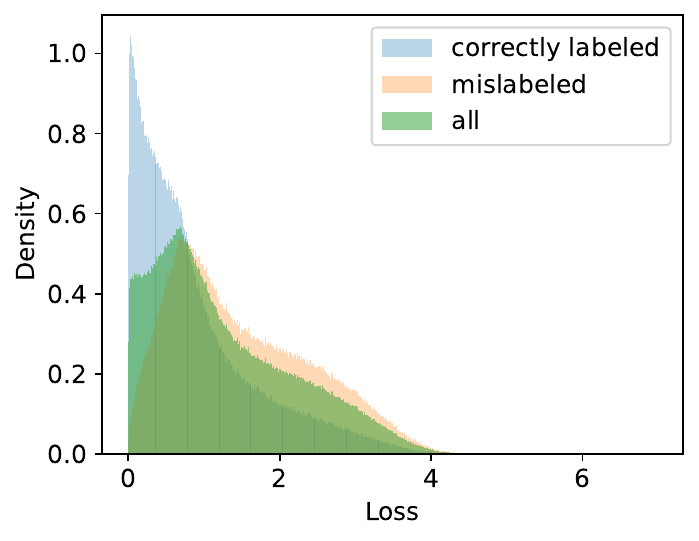}}
  \subfloat[Initial Model, Log Scale]{
   \includegraphics[width=0.245\linewidth]{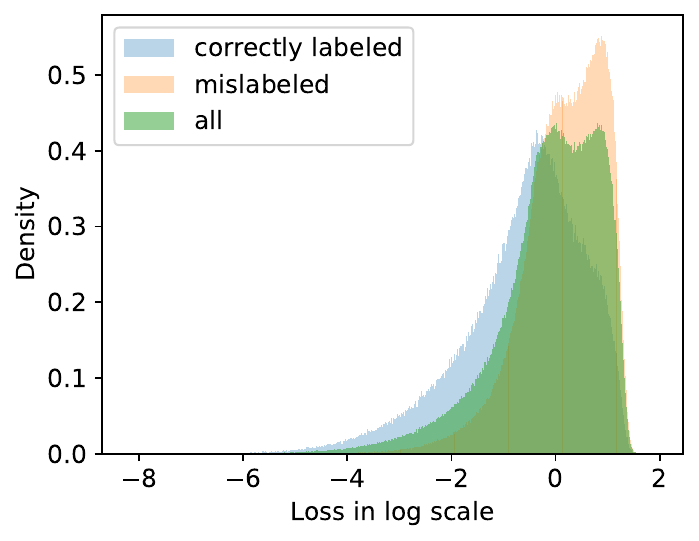}}
  \subfloat[Reflective Learning, Linear Scale]{
   \includegraphics[width=0.245\linewidth]{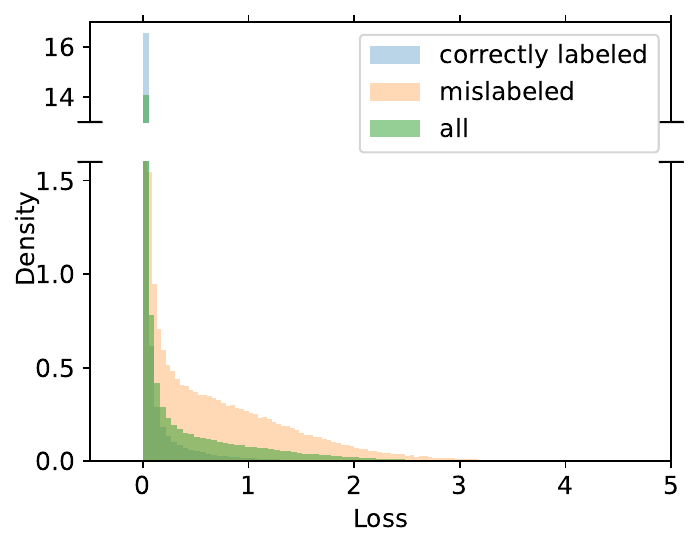}}
  \subfloat[Reflective Learning, Log Scale]{
   \includegraphics[width=0.245\linewidth]{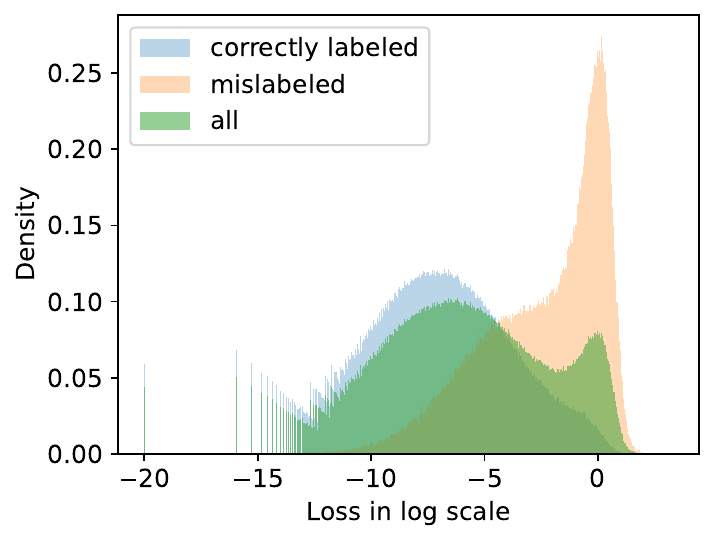}}
  \caption{Histogram of the cross entropy loss between the model's prediction and the pseudo label. (a) and (b) are produced by the model trained with the initial pseudo label without applying the proposed SSRL approach. (c) and (d) are produced by the teacher model after 30 epochs of SSRL.}
  \label{fig: training loss}
\end{figure*}

\subsection{Online clustering}
\label{sec: online clustering}
In the cluster assignment task, the objective is to maximize the alignment of the cluster assignments $q(k|\x_i)$ with the predicted class probabilities $p_t(k|\x_i)$ provided by a teacher model, ensuring that each data point is assigned to the cluster where it best fits according to these predictions.
\begin{equation}
\label{eq: online clustering}
\begin{aligned}
	& \max_{q} \frac{1}{N} \sum_{i=1}^{N} \sum_{k=1}^{K} q(k|\x_i)  p_t(k|\x_i)\\
	\text{subject to } & \forall k: q(k|\x_i)\in \{0, 1\} \text{ and} \sum_{k=1}^{K} q(k|\x_i) = 1
\end{aligned}
\end{equation}
To address this, we explore two online clustering methodologies:

\subsubsection{Direct maximum probability assignment}
The most intuitive method generates cluster assignment based on the highest predicted probability class from the teacher:
\begin{equation}
q(k|\x_i) = \delta\left(k - \arg\max_j p_t(j|\x_i)\right)
\end{equation}
Here, $\delta(k - \arg\max_j p_t(j | x_i))$ is the Kronecker delta function, defined as 1 when  $k = \arg\max_j p_t(j | x_i)$ and 0 otherwise.
Essentially, each data sample is allocated to the cluster corresponding to the class that the teacher model is most confident in.

\subsubsection{Cluster assignment through optimal transport}
Drawing inspiration from SeLa \cite{asano_self-labelling_2020}, we introduce an added constraint to the objective in Equation \ref{eq: online clustering}, ensuring that the $N$ training samples are distributed evenly across the $K$ clusters:
\begin{equation}
\label{eq: constrained online clustering}
\begin{aligned}
    & \max_{q} \frac{1}{N} \sum_{i=1}^{N} \sum_{k=1}^{K} q(k|\x_i) p_t(k|\x_i) \\
    \text{subject to } & \forall k: q(k|\x_i)\in \{0, 1\},  \sum_{i=1}^{N} q(k|\x_i) = \frac{N}{K}
\end{aligned}
\end{equation}
Such constraints ensure a distinct label for every data point and a uniform distribution of the $N$ samples over the $K$ classes, preventing identical pseudo labeling for all training samples.

Building on the perspective of SeLa \cite{asano_self-labelling_2020}, the optimization problem depicted in Equation \ref{eq: constrained online clustering} can be mapped to an optimal transport problem \cite{cuturi_sinkhorn_2013}.
To understand this, let's define $P$ as the $K \times N$ matrix where $P_{ki}=\frac{1}{N}p_t(k|\x_i)$, and $Q$ as the $K \times N$ matrix of assigned joint probabilities between $a$ and $b$ with $Q_{ki}=\frac{1}{N}q(k|\x_i)$.
Following the notation in \cite{cuturi_sinkhorn_2013}, $Q$ can be conceptualized as an element of the transportation polytope:
\begin{equation}
	U(\r,\cc) := \left\{ Q \in \mathbb{R}_{+}^{K\times N} | Q\mathds{1}=\r, Q^T\mathds{1}=\cc \right\}
\end{equation}
where $\mathds{1}$ is the vector of all ones of appropriate dimension.
Based on the given constraints, we get:
\begin{equation}
	\r = \frac{1}{K}\cdot\mathds{1}; \ \ \ \ \cc = \frac{1}{N}\cdot\mathds{1}
\end{equation}
Given matrices $P$ and $Q$, the objective function in Equation \ref{eq: constrained online clustering} can be recast as:
\begin{equation}
	\frac{1}{N} \sum_{i=1}^{N} \sum_{k=1}^{K} q(k|\x_i) p_t(k|\x_i) = \langle Q, P \rangle 
\end{equation}
where $\langle\cdot\rangle$ is the Frobenius dot-product between two matrices.
Consequently, Equation \ref{eq: constrained online clustering} can be translated into an optimal transport problem between $\r$ and $\cc$ with a cost of $-P$:
\begin{equation}
	\label{eq: ot}
	\min_{Q\in U(\r,\cc) } \langle Q, -P \rangle
\end{equation}

To expedite the optimal transport solver, an entropic constraint was integrated into the classical optimal transport problem as introduced by Cuturi \cite{cuturi_sinkhorn_2013}.
This regularization of the problem is defined by:
\begin{equation}
	U_\alpha(\r,\cc) := \left\{ Q \in U(\r,\cc) \ | \ \mathrm{KL}(Q\|\r\cc^T) \leq \alpha \right\}
\end{equation}
where KL represents the Kullback-Leibler divergence.
Given the concavity of entropy, we have $U_\alpha(\r,\cc) \subset U(\r,\cc)$.
Consequently, the optimal transport problem (as shown in Equation \ref{eq: ot}) is reframed as:
\begin{equation}
	\min_{Q\in U_\alpha(\r,\cc) } \langle Q, - P \rangle
\end{equation}
Introducing a Lagrange multiplier for the entropy constraint, we arrive at the dual optimization problem:
\begin{equation}
\label{eq: ot reg}
	\min_{Q\in U(\r,\cc) } \langle Q, - P \rangle + \frac{1}{\lambda} \mathrm{KL}(Q\|\r\cc^T)
\end{equation}
From the Lagrangian of Equation \ref{eq: ot reg}, we can express the minimizer of Equation \ref{eq: ot reg} as:
\begin{equation}
	Q = \mathrm{diag}(\u)e^{\lambda P}\mathrm{diag}(\vv)
\end{equation}
In the equation, exponentiation is carried out element-wise. Additionally, $\u$ and $\vv$ are two non-negative vectors that serve as scaling coefficients, ensuring the resulting matrix $Q$ adheres to the probability matrix standards.

The Sinkhorn-Knopp algorithm is employed to determine the optimal $Q$. This algorithm iteratively adjusts the rows and columns of the matrix utilizing diagonal matrices until a convergence point is reached:
\begin{equation}
	\forall k: \u_k \leftarrow \frac{\r_k}{[e^{\lambda P} \vv]_k}; \ \ \ \ \forall i: \vv_i \leftarrow  \frac{\cc_i}{[\u^T e^{\lambda P}]_i}
\end{equation}

To generate pseudo labels using the Sinkhorn-Knopp algorithm, we employ a batched approach.
Specifically, we accumulate the matrix $P$ over $M$ batches with batch size $B$, ensuring total number of training samples $N=M\times B$ is larger than the number of cluster $K$.
Every $M$ batches, we update the cluster assignments utilizing the Sinkhorn-Knopp algorithm. 
This method provides a computationally efficient way to handle large datasets, ensuring consistent and optimized pseudo-label assignments in line with the teacher's predictions.

Either with direct maximum probability assignment or optimal transport-based cluster assignment, the assigned clusters are determined based on the teacher model’s predictions, without constraints ensuring that every output class will receive data samples. As training progresses, clusters with extremely low prediction confidence shrink and eventually disappear. This phenomenon is observed in our experiments, as described later in Section \ref{sec: labeling} and Figure \ref{fig:labelingvsepoch}, where the number of active clusters gradually decreases during training until a stable count is reached.

SSRL naturally maintains stable and meaningful cluster assignments throughout training. Starting from initial pseudo labels, the online clustering continuously refines assignments as the teacher model's discrimination ability improves. When a sample's current assignment becomes suboptimal, the teacher model reassigns it based on learned representations. Online clustering works organically with EMA updates - EMA ensures smooth model evolution while preserving previous knowledge, enabling stable and consistent refinements to pseudo labels.

\subsection{Pseudo label correction}
To further refine the pseudo label generation process, we introduce a label correction mechanism employing a pseudo label queue.
This queue retains a history of pseudo labels previously generated by the teacher model for each training sample.
With a predetermined fixed length $L$, the queue ensures consideration only of the most recent $L$ predictions.
To filter out sporadic or outlier predictions and cultivate a robust pseudo label, we employ a statistical mode evaluation of the labels within the queue.
This ensures that the most frequently occurring label in the recent history is selected as the final pseudo label, thereby enhancing the reliability of the label assignment and mitigating the effects of transient erroneous predictions.

\subsection{Noisy label modeling}
To mitigate the challenges posed by noisy pseudo labels, our framework incorporates a strategy to model label noise following the approach in \cite{arazo_unsupervised_2019}. Prior research \cite{arazo_unsupervised_2019} has shown that deep neural networks (DNNs) tend to learn correctly labeled samples first before gradually fitting to mislabeled ones. As a result, mislabeled samples typically exhibit higher loss values compared to correctly labeled ones, allowing us to leverage this property for noise modeling.

Figure \ref{fig: training loss} shows an illustration of such behavior.
Since the pseudo label is estimated in the self-supervised setting, we do not have the ground truth references for the correct and incorrect labels.
To estimate this noisy label information, we employ the Hungarian algorithm, mapping the pseudo labels to the ground truth labels.
Figure \ref{fig: training loss} exhibit a bimodal distribution with two distinct peaks of the logarithmically scaled losses.
By modeling this loss distribution, we can effectively segregate accurately labeled data from the mislabeled, which then aids in computing cleaner label probabilities for the training set.

To achieve this, we use a two-component GMM to model the logarithmically scaled losses generated by the teacher model.
Mathematically, the mixture model can be expressed as:
\begin{equation}\begin{aligned}
	p(\ell_t) = \pi & \cdot \mathcal{N}(\log(\ell_t); \mu_1, \sigma_1^2) +\\
	(1 - \pi) & \cdot \mathcal{N}(\log(\ell_t); \mu_2, \sigma_2^2)
\end{aligned}\end{equation}
For sample $\x_i$, $\ell_{t,i}$ represents the cross entropy loss between the teacher's prediction and its pseudo label. The term $p(\ell_t)$ represents the probability distribution of $\log(\ell_t)$. The coefficient $\pi$ is the mixture weight, and $\mathcal{N}(\log(\ell_t); \mu, \sigma^2)$ is the Gaussian distribution parameterized by mean $\mu$ and variance $\sigma^2$.

The GMM aids in distinguishing between the loss distributions of clean labels and those of noisy labels.
After establishing this loss distribution model, a clean label probability is assigned to each training sample as:
\begin{equation}
p_{\text{clean}}(\ell_t) = \frac{\pi \cdot \mathcal{N}(\log(\ell_t); \mu_1, \sigma_1^2)}{p(\ell_t)}
\end{equation}
Given that samples with clean labels yield lower losses, the Gaussian component $\mathcal{N}(\log(\ell_t); \mu_1, \sigma_1^2)$ associated with these samples has a smaller mean, i.e., $\mu_1 < \mu_2$.
Utilizing the clean label probability, $p_{\text{clean}}(\ell_t)$, the final loss is adjusted, directing the model to give greater emphasis to samples deemed to have accurate labels:
\begin{equation}
	\mathcal{L} = -\frac{1}{N}\sum_{i=1}^{N}p_{\text{clean}}(\ell_{t,i}) \log p_s(y_i|\x_i')
\end{equation}

Combining all the methods discussed above, Algorithm \ref{algo: ssrl} presents the complete procedure of the proposed SSRL method. Here, we apply the direct maximum probability assignment as the online clustering method. It can easily be extended to the optimal transport method.

\section{Experimental setups}

\subsection{Data}
The experiments are conducted on the VoxCeleb dataset \cite{nagrani_voxceleb:_2017, chung_voxceleb2:_2018}. For model training, we use the development set of VoxCeleb 2, which contains 1,092,009 audio files from 5,994 speakers. While speaker identity labels are available, they are only used for experimental analysis and not for model training.

For evaluation, we report speaker verification results using three trial lists from the VoxCeleb 1 dataset as defined in \cite{chung_voxceleb2:_2018}:
\begin{itemize}
\item VoxCeleb 1-O: The original trial list with 37,720 trials from 40 speakers.
\item VoxCeleb 1-E: An extended trial list with 581,480 trials from 1,251 speakers.
\item VoxCeleb 1-H: A hard trial list with 552,536 trials from 1,190 speakers, where all test pairs share the same language and gender.
\end{itemize}

\subsection{Data Augmentation}
Data augmentation is effective for deep speaker representation learning in both supervised learning \cite{cai_within-sample_2020} and contrastive self-supervised learning \cite{inoue_semi-supervised_2020, huh_augmentation_2020, chen_simple_2020}. We utilized two primary strategies:
\begin{itemize}
\item Additive noise augmentation: The MUSAN dataset \cite{musan} was used as our noise source, adding ambient noise, musical sounds, and babble noise to our audio files. Babble noise was generated by merging three to eight separate speech files from the MUSAN dataset, with signal-to-noise ratios (SNR) ranging from 0 to 20 dB.
\item Convolutional reverberation noise augmentation: We used 40,000 simulated room impulse responses (RIR) from small to medium-sized rooms, as described in \cite{ko2017study}.
\end{itemize}
To maintain variability during training, we applied on-the-fly data augmentation. In SSRL training, the student network was trained with two-thirds of the data augmented utterances, while the teacher network used unaltered speech data.

\subsection{Implementation details}
We evaluate the proposed methods on two different network architectures for speaker representation learning: ResNet \cite{he_deep_2016} and ECAPA-TDNN \cite{desplanques_ecapa-tdnn_2020}. The baseline method used for comparison is the two-stage iterative framework. For each network architecture, a supervised model is trained to serve as a reference point (upper bound) for model performance, using the same training hyperparameters as those in the second stage of the two-stage iterative framework.

\subsubsection{ResNet -- two-stage iterative framework}
We first use the two-stage iterative framework trained on ResNet \cite{he_deep_2016} as the baseline, following our previous research on the two-stage iterative framework \cite{cai_iterative_2021, 9741340}.

In the first stage, we apply contrastive self-supervised learning (CSL) \cite{chen_simple_2020} to learn speaker representations. In the second stage (iterative training), initial pseudo labels for the training dataset are generated using K-means clustering on speaker embeddings from CSL. The number of clusters is set to 6,000, the same as in \cite{9741340}, where it was determined using the elbow method. For network architecture, hyperparameters, and other training details, readers can refer to \cite{9741340}.

\subsubsection{ResNet -- improved two-stage framework with SSRL}
\label{sec: resnet setup}
For the improved two-stage framework with SSRL trained on ResNet, the first stage remains the same as in the two-stage iterative framework. To initiate second-stage training, the number of clusters for K-means is set to 8,000, which is higher than the 6,000 clusters used in the two-stage iterative framework. This adjustment is made for two reasons: (1) The elbow method identifies a reasonable cluster count between 5,000 and 8,000 \cite{9741340}. (2) As discussed in Section \ref{sec: online clustering}, the cluster count naturally decreases due to the online clustering process. Setting a higher initial cluster count ensures sufficient granularity, allowing the model to refine pseudo labels without collapsing clusters too early.

In the second stage, to initialize SSRL training, the ResNet-based speaker embedding network is trained for 55 epochs with initial pseudo labels.
A cosine annealing scheduler adjusts the learning rate from 1e-3 to 1e-5, including a 5-epoch warm-up phase. The batch size is set to 512, and the Adam optimizer is applied.

During the SSRL training phase, audio waveforms are cropped to 2 seconds for the student model and 6 seconds for the teacher model.
The student network is trained for 100 epochs using the Adam optimizer, with the learning rate scheduled via cosine annealing from 5e-4 to 1e-5.
The loss function used is cross entropy.
The pseudo label queue length is set to 5 unless stated otherwise.
The EMA momentum parameter, denoted as $\lambda$ in Equation \ref{eq: ema}, linearly increases from 0.999 to 0.9999 during SSRL training.

\subsubsection{ECAPA-TDNN -- two-stage iterative framework}
To compare with other studies, we also adopt the ECAPA-TDNN-based speaker embedding network \cite{desplanques_ecapa-tdnn_2020} as an alternative backbone.

For the first stage self-supervised training, ECAPA-TDNN speaker embedding network \cite{desplanques_ecapa-tdnn_2020} is pretrained with DINO \cite{caron_emerging_2021}. Following the structure in \cite{chen_pushing_2023}, the ECAPA-TDNN network has channels sequenced as 1024, 1024, 1024, 1024, and 3072 across the initial TDNN layer and four TDNN blocks. After the ECAPA-TDNN encoder, we use attentive statistical pooling followed by a 512-dimensional fully connected layer for speaker embeddings. The DINO projection head includes four fully connected layers with hidden dimensions of 2048, 2048, 8192, and 256, ending with a 65536-dimensional weight-normalized fully connected layer. We employ multi-crop data augmentation, giving the EMA teacher two 4-second data-augmented views and the student four 2-second data-augmented views for each training sample.

The DINO pretraining uses a stochastic gradient descent (SGD) optimizer over 100 epochs, with a cosine annealing scheduler modulating the learning rate from 0.2 to 1e-5, including a 10-epoch warm-up phase. The temperature hyperparameters for cross-entropy are set to 0.04 for the teacher and 0.1 for the student. For more detailed training procedures, refer to \cite{caron_emerging_2021} and \cite{chen_pushing_2023}.

The comparison of different first stage models used in this work can be found in Table \ref{tab: csl and dino}. Unlike random initialization, stage 1 provides a structured representation for clustering, enabling the first clustering round to generate more reliable pseudo labels. This improves the quality of subsequent second stage training, ensuring the model refines meaningful speaker representations rather than noise.

\begin{table}[t]
  \caption{Comparison of two self-supervised pretrained models. EER is evaluated on VoxCeleb 1-O; labeling metrics are based on k-means clustering with 8,000 clusters.}
  \label{tab: csl and dino}
  \centering
  \begin{tabular}[c]{@{\ }l@{\ }l@{\ }c@{\ }c@{\ }c@{\ }c@{\ }c@{\ }}
    \toprule
    \tabincell{l}{Pretrained\\Method} & \tabincell{l}{Network\\Architecture} & \#Parameters & EER\textdownarrow & NMI \textuparrow& Accuracy \textuparrow& Purity\textuparrow \\
    \midrule
    CSL & ResNet & 1.37M & 8.86\% & 0.7744 & 36.87\% & 55.32\% \\
    \midrule
    DINO & ECAPA-TDNN & 63.65M\tablefootnote{The ECAPA-TDNN encoder has a total of 22.73 million parameters. The DINO projection head contains 40.92 million parameters. The projection head is only used during DINO training; speaker embeddings are extracted from the output of the ECAPA-TDNN encoder.} & 2.94\% & 0.9319 & 65.26\% & 88.52\% \\
    \bottomrule
  \end{tabular}
\end{table}

In the second stage of iterative training, pseudo labels are generated by applying K-means clustering to the speaker embeddings from the previous training round, targeting 8,000 clusters. Each training round employs the Adam optimizer with a batch size of 480, and the learning rate is managed by a cosine annealing scheduler, transitioning from 1e-4 to 1e-5 over 40 epochs. 

To ensure stable training, we initialize the encoder's parameters with DINO pre-trained parameters for the first training round. In subsequent training rounds, the encoder retains the parameters from the previous training round. The predictor, i.e., the final linear layer for speaker classification, is reinitialized using K-means cluster centers.

\subsubsection{ECAPA-TDNN -- improved two-stage framework with SSRL}
For the improved two-stage framework with SSRL trained on ECAPA-TDNN, the first stage remains the same as in the two-stage iterative framework.

In the second stage, we directly initialize both the student and teacher networks using DINO-pretrained parameters and apply SSRL. The predictor, a single linear layer, has its weights initialized with the 8,000 K-means cluster centers, while the biases are set to zero. The training batch size for the ECAPA-TDNN model is 480, and other training configurations for SSRL remain the same as those for the ResNet-based pipeline.
For the training objective, in addition to cross-entropy loss, we train another ECAPA-TDNN with SSRL using the additive angular margin (AAM) loss \cite{deng_arcface_2019} to further enhance the model’s capacity. The AAM loss margin is set to 0.2, and the scaling factor is 32.

\subsection{Evaluation metric}
\subsubsection{Speaker verification evaluation}

We assess the effectiveness of speaker verification systems by measuring the equal error rate (EER) and the minimum detection cost (minDCF) \cite{nist_nist_2016}. For the detection cost function, we configure the parameters as $C_{\text{Miss}}=1$, $C_{\text{FA}}=1$, and $P_{\text{Target}}=0.05$.

\subsubsection{Clustering evaluation}

To evaluate clustering quality, we use three metrics as outlined in \cite{asano_labelling_2020} and \cite{9741340}:
\begin{itemize}
	\item Normalized mutual information (NMI): This metric measures the agreement between our clustering and the true data grouping, providing a score between 0 and 1, where 0 indicates no match and 1 indicates a perfect match.
	\item Clustering accuracy: We evaluate accuracy by comparing pseudo labels to ground truth labels, using the Hungarian algorithm \cite{munkres1957algorithms} to establish label correspondence.
	\item Mean maximal purity per cluster: This metric assesses the semantic purity of each pseudo cluster in comparison to the ground truth labels:
	\begin{equation}
	\mathrm{purity} = \frac{1}{K} \sum_{k\in K} \max\left(p\left( y | \hat{y}=k \right) \right)
\end{equation}
where $K$ is the number of pseudo clusters, $\hat{y}$ represents a pseudo cluster and $p\left( y | \hat{y}=k \right)$ is the distribution of ground-truth labels within pseudo cluster $k$.
\end{itemize}

\section{Experimental results}
This section evaluates the improved two-stage framework with SSRL in terms of speaker verification performance and pseudo-labeling robustness.
We also investigates the contributions of different individual components in the proposed SSRL method.

\subsection{Speaker verification performance}

\begin{table}[t!]
  \caption{ResNet results: speaker verification performance (minDCF and EER[\%]) on VoxCeleb 1 test trials.}
  \label{tab: ssrl result1}
  \centering
  \begin{tabular}[c]{@{\ }l@{\ }l@{\ }c@{\ }c@{\ }c@{\ }c@{\ }c@{\ }c@{\ }}
    \toprule
    \multirow{2}*{\textbf{Model}} & & \multicolumn{2}{@{}c@{}}{\textbf{VoxCeleb 1-O}} & \multicolumn{2}{@{}c@{}}{\textbf{VoxCeleb 1-E}} & \multicolumn{2}{@{}c@{}}{\textbf{VoxCeleb 1-H}} \\
    \cmidrule(lr){3-4} \cmidrule(lr){5-6} \cmidrule(lr){7-8}
    & & minDCF & EER & minDCF & EER & minDCF & EER \\
    \midrule
    \multicolumn{2}{l}{Supervised} & 0.097 & 1.51 & 0.102 & 1.59 & 0.178 & 3.00 \\
    \midrule
    \multirow{6}*{\tabincell{l}{Two-\\Stage\\Iterative\\Framework\\\cite{9741340}}}
    & CSL (Stage 1) & 0.508 & 8.86 & 0.570 & 10.15 & 0.710 & 16.20 \\
    & Round 1 & 0.257 & 3.64 & 0.299 & 4.11 & 0.459 & 7.68 \\
    & Round 2 & 0.214 & 2.99 & 0.234 & 3.41 & 0.362 & 6.25 \\
    & Round 3 & 0.190 & 2.93 & 0.214 & 3.23 & 0.334 & 5.85 \\
    & Round 4 & 0.184 & 2.85 & 0.202 & 3.16 & 0.314 & 5.54 \\
    & Round 5 & 0.173 & 2.74 & 0.201 & 3.08 & 0.311 & 5.48\\
    \midrule
    \multicolumn{2}{l}{SSRL (one round)} & \textbf{0.163} & \textbf{2.39} & \textbf{0.183} & \textbf{2.63} & \textbf{0.285} & \textbf{4.74} \\
    \bottomrule
  \end{tabular}
\end{table}

\begin{table}[t!]
  \caption{ECAPA-TDNN results: Speaker verification performance (minDCF and EER[\%]) on VoxCeleb 1 test trials.}
  \label{tab: ssrl result2}
  \centering
  \begin{tabular}[c]{@{\ }l@{\ }l@{\ }c@{\ }c@{\ }c@{\ }c@{\ }c@{\ }c@{\ }}
    \toprule
    \multirow{2}*{\textbf{Model}} & & \multicolumn{2}{@{}c@{}}{\textbf{VoxCeleb 1-O}} & \multicolumn{2}{@{}c@{}}{\textbf{VoxCeleb 1-E}} & \multicolumn{2}{@{}c@{}}{\textbf{VoxCeleb 1-H}} \\
    \cmidrule(lr){3-4} \cmidrule(lr){5-6} \cmidrule(lr){7-8}
    & & minDCF & EER & minDCF & EER & minDCF & EER \\
    \midrule
    \multicolumn{2}{l}{Supervised} & 0.143 & 1.88 & 0.136 & 1.98 & 0.237 & 3.96 \\
    \multicolumn{2}{l}{Supervised + AAM} & 0.075 & 0.99 & 0.081 & 1.22 & 0.144 & 2.35 \\
    \midrule
    \multirow{4}*{\tabincell{l}{Two-\\Stage\\Iterative\\Framework}}
    & DINO (Stage 1) & 0.202 & 2.94 & 0.218 & 3.05 & 0.364 & 5.88 \\
    & Round 1 & 0.181 & 2.49 & 0.183 & 2.73 & 0.288 & 5.01 \\
    & Round 2 & 0.174 & 2.34 & 0.180 & 2.66 & 0.282 & 4.90 \\
    & Round 3 & 0.177 & 2.28 & 0.184 & 2.70 & 0.288 & 4.95 \\
    \midrule
    \multicolumn{2}{l}{SSRL (one round)} & 0.131 & 1.77 & 0.127 & 1.85 & 0.217 & 3.59 \\
    \multicolumn{2}{l}{SSRL (one round) + AAM} & \textbf{0.101} & \textbf{1.25} & \textbf{0.098} & \textbf{1.47} & \textbf{0.174} & \textbf{2.86} \\
    \bottomrule
  \end{tabular}
\end{table}

\subsubsection{Comparing SSRL with iterative training in two-stage framework}
The primary objective of our experiments is to compare the proposed SSRL method with the iterative training in the two-stage framework.
Table \ref{tab: ssrl result1} shows that the SSRL-trained ResNet model achieves an EER of 2.39\% on the VoxCeleb 1-O trial in just one training round, surpassing the fifth-round model in the two-stage iterative framework (2.74\%).

Similarly, for the ECAPA-TDNN model in Table \ref{tab: ssrl result2}, the SSRL method demonstrates superior performance with an EER of 1.77\% in one training round, compared to 2.28\% EER from the third-round model in the two-stage iterative framework. The integration of AAM loss in SSRL suggests even more potential, with EER dropping to 1.25\%.

The supervised results in Tables \ref{tab: ssrl result1} and \ref{tab: ssrl result2} serve as upper bounds for model performance. Compared to the supervised model, both self-supervised methods do not surpass supervised performance. However, SSRL significantly reduces the performance gap. For the ResNet-based pipeline, SSRL achieves an EER of 2.39\% on VoxCeleb 1-O, compared to 2.74\% for the best model in two-stage iterative framework. For ECAPA-TDNN, SSRL achieves 1.77\% EER, improving upon the two-stage iterative framework’s best result of 2.28\%, bringing it closer to the supervised model’s 1.88\% EER.

The superiority of the SRRL second stage over the iterative second stage can be ascribed to its robust pseudo-labeling mechanism.
Unlike the two-stage iterative framework which employs static pseudo labels for a whole training round, SSRL benefits from dynamically updated labels via self-supervised knowledge distillation and online clustering.
This continuous refinement ensures the student model always benefits from the latest supervision signals, eliminating the `stale' label problem observed in the two-stage iterative framework.
Furthermore, SSRL's incorporation of a pseudo label queue and noisy label modeling techniques further improve the reliability and robustness of the pseudo labels, enhancing overall model performance.

\begin{figure}[t!]
     \centering
    \subfloat[ResNet-based pipeline]{
    	\includegraphics[width=0.65\linewidth]{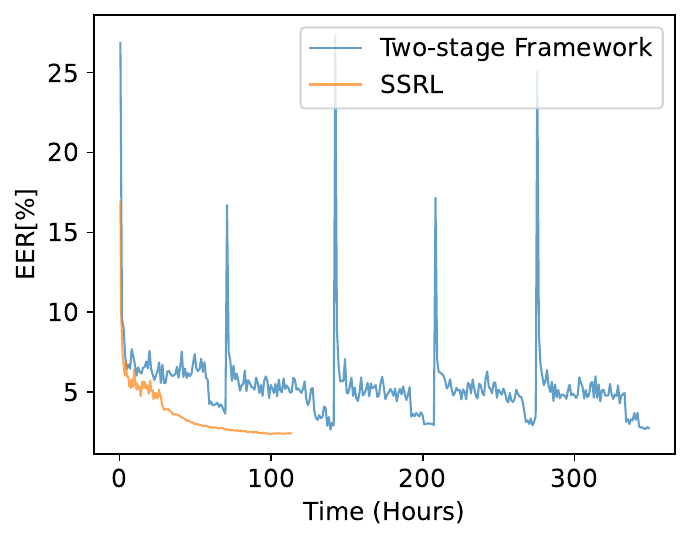}}
\\    \subfloat[ECAPA-TDNN-based pipeline]{
    	\includegraphics[width=0.65\linewidth]{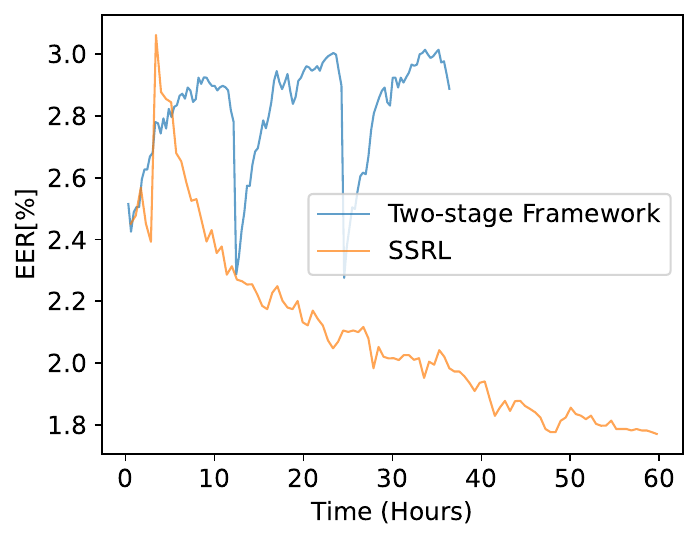}}
     \caption{Comparison of training time vs. EER for iterative training and SSRL training in the two-stage framweork}
  \label{fig:eervstime}
\end{figure}

\begin{figure*}[t]
    \centering
\begin{tikzpicture}
    \begin{groupplot}[
        group style={
            group size=4 by 1,
            horizontal sep=0.068\textwidth,
        },
        ymajorgrids=true,
        grid style=dashed,
        width=0.183\textwidth,
        height=3.3cm,
        scale only axis,
        x label style={font=\footnotesize},
        y label style={font=\footnotesize, at={(axis description cs:0.09,0.5)}},
        title style={font=\footnotesize},
        tick label style={font=\footnotesize},
        legend style={transpose legend, legend columns=0, font=\footnotesize},
    ]
        
        \nextgroupplot[xlabel={SSRL Training Epoch}, ylabel={Number of Clusters}, xtick={0, 50, 100}, xticklabels={0, 50, 100}, ytick={5084, 5328, 8000}, yticklabels={5084, 5328, 8000}, legend to name=grouplegend]
            \addplot+[myBlue, line width=0.6pt, mark=none] coordinates {(0, 8000) (1, 7978) (2, 7957) (3, 7917) (4, 7690) (5, 7233) (6, 6917) (7, 6696) (8, 6444) (9, 6221) (10, 6042) (11, 5919) (12, 5818) (13, 5738) (14, 5666) (15, 5619) (16, 5569) (17, 5527) (18, 5486) (19, 5460) (20, 5426) (21, 5408) (22, 5385) (23, 5359) (24, 5341) (25, 5330) (26, 5314) (27, 5296) (28, 5290) (29, 5281) (30, 5266) (31, 5251) (32, 5242) (33, 5238) (34, 5230) (35, 5220) (36, 5213) (37, 5204) (38, 5201) (39, 5197) (40, 5191) (41, 5184) (42, 5178) (43, 5172) (44, 5169) (45, 5165) (46, 5162) (47, 5161) (48, 5159) (49, 5156) (50, 5153) (51, 5149) (52, 5147) (53, 5146) (54, 5139) (55, 5135) (56, 5132) (57, 5127) (58, 5125) (59, 5123) (60, 5120) (61, 5120) (62, 5114) (63, 5111) (64, 5110) (65, 5109) (66, 5105) (67, 5104) (68, 5103) (69, 5102) (70, 5099) (71, 5097) (72, 5095) (73, 5094) (74, 5092) (75, 5091) (76, 5091) (77, 5090) (78, 5090) (79, 5090) (80, 5089) (81, 5089) (82, 5088) (83, 5086) (84, 5085) (85, 5085) (86, 5085) (87, 5085) (88, 5085) (89, 5085) (90, 5085) (91, 5085) (92, 5085) (93, 5085) (94, 5085) (95, 5085) (96, 5085) (97, 5085) (98, 5084) (99, 5084) (100, 5084)};
          \addplot+[myRed, line width=0.6pt, mark=none] coordinates{(0,8000)(1,8000)(2,7997)(3,7995)(4,7946)(5,7742)(6,7482)(7,7265)(8,7087)(9,6928)(10,6781)(11,6670)(12,6572)(13,6471)(14,6377)(15,6296)(16,6220)(17,6150)(18,6077)(19,6011)(20,5967)(21,5918)(22,5875)(23,5842)(24,5819)(25,5800)(26,5773)(27,5755)(28,5734)(29,5718)(30,5707)(31,5703)(32,5695)(33,5688)(34,5681)(35,5676)(36,5674)(37,5668)(38,5660)(39,5657)(40,5653)(41,5651)(42,5648)(43,5646)(44,5642)(45,5640)(46,5639)(47,5639)(48,5638)(49,5637)(50,5636)(51,5628)(52,5614)(53,5601)(54,5576)(55,5551)(56,5527)(57,5499)(58,5476)(59,5451)(60,5435)(61,5421)(62,5403)(63,5390)(64,5379)(65,5371)(66,5366)(67,5360)(68,5352)(69,5346)(70,5341)(71,5337)(72,5335)(73,5334)(74,5334)(75,5332)(76,5332)(77,5331)(78,5330)(79,5330)(80,5330)(81,5330)(82,5330)(83,5330)(84,5330)(85,5330)(86,5330)(87,5330)(88,5330)(89,5330)(90,5329)(91,5328)(92,5328)(93,5328)(94,5328)(95,5328)(96,5328)(97,5328)(98,5328)(99,5328)(100,5328)};
            \addlegendentry{SSRL-trained ResNet};
            \addlegendimage{myRed, line width=0.6t}
            \addlegendentry{SSRL-trained ECAPA-TDNN}
            
        \nextgroupplot[xlabel={SSRL Training Epoch}, ylabel={NMI}, xtick={0, 50, 100}, xticklabels={0, 50, 100}, ytick={0.8597, 0.9337, 0.9651}, yticklabels={0.860, 0.934, 0.965}]
            \addplot+[myBlue, line width=0.6pt, mark=none] coordinates {(0, 0.8597) (1, 0.8684) (2, 0.8681) (3, 0.8699) (4, 0.871) (5, 0.8719) (6, 0.8739) (7, 0.8769) (8, 0.88) (9, 0.8828) (10, 0.8855) (11, 0.8879) (12, 0.8901) (13, 0.8923) (14, 0.8943) (15, 0.8962) (16, 0.898) (17, 0.8996) (18, 0.9011) (19, 0.9025) (20, 0.9038) (21, 0.905) (22, 0.9062) (23, 0.9073) (24, 0.9083) (25, 0.9093) (26, 0.9101) (27, 0.911) (28, 0.9118) (29, 0.9127) (30, 0.9134) (31, 0.9143) (32, 0.915) (33, 0.9156) (34, 0.9163) (35, 0.9169) (36, 0.9176) (37, 0.9181) (38, 0.9186) (39, 0.9192) (40, 0.9197) (41, 0.9202) (42, 0.9207) (43, 0.9212) (44, 0.9217) (45, 0.9222) (46, 0.9225) (47, 0.923) (48, 0.9234) (49, 0.9238) (50, 0.9242) (51, 0.9246) (52, 0.9249) (53, 0.9253) (54, 0.9257) (55, 0.9261) (56, 0.9263) (57, 0.9266) (58, 0.9269) (59, 0.9272) (60, 0.9275) (61, 0.9278) (62, 0.9282) (63, 0.9284) (64, 0.9287) (65, 0.929) (66, 0.9292) (67, 0.9294) (68, 0.9297) (69, 0.9299) (70, 0.9301) (71, 0.9303) (72, 0.9306) (73, 0.9308) (74, 0.931) (75, 0.9311) (76, 0.9313) (77, 0.9314) (78, 0.9316) (79, 0.9317) (80, 0.9319) (81, 0.9321) (82, 0.9322) (83, 0.9323) (84, 0.9324) (85, 0.9325) (86, 0.9327) (87, 0.9327) (88, 0.9328) (89, 0.933) (90, 0.933) (91, 0.933) (92, 0.9332) (93, 0.9332) (94, 0.9333) (95, 0.9334) (96, 0.9334) (97, 0.9335) (98, 0.9336) (99, 0.9336) (100, 0.9337)};
          \addplot+[myRed, line width=0.6pt, mark=none] coordinates{(0,0.9352)(1,0.9365)(2,0.9376)(3,0.9387)(4,0.9396)(5,0.9398)(6,0.9407)(7,0.9421)(8,0.9434)(9,0.9444)(10,0.9452)(11,0.9459)(12,0.9466)(13,0.9472)(14,0.9478)(15,0.9483)(16,0.9489)(17,0.9494)(18,0.9499)(19,0.9505)(20,0.951)(21,0.9516)(22,0.952)(23,0.9525)(24,0.9531)(25,0.9535)(26,0.954)(27,0.9544)(28,0.9549)(29,0.9553)(30,0.9557)(31,0.9561)(32,0.9565)(33,0.9569)(34,0.9572)(35,0.9575)(36,0.9578)(37,0.958)(38,0.9583)(39,0.9586)(40,0.9588)(41,0.9591)(42,0.9594)(43,0.9596)(44,0.9598)(45,0.9601)(46,0.9602)(47,0.9604)(48,0.9605)(49,0.9604)(50,0.9603)(51,0.9601)(52,0.9598)(53,0.9597)(54,0.9596)(55,0.9595)(56,0.9595)(57,0.9595)(58,0.9595)(59,0.9596)(60,0.9596)(61,0.9598)(62,0.9599)(63,0.9601)(64,0.9603)(65,0.9605)(66,0.9607)(67,0.961)(68,0.9613)(69,0.9614)(70,0.9616)(71,0.9618)(72,0.962)(73,0.9622)(74,0.9624)(75,0.9626)(76,0.9627)(77,0.9629)(78,0.963)(79,0.9632)(80,0.9633)(81,0.9634)(82,0.9636)(83,0.9637)(84,0.9638)(85,0.964)(86,0.9641)(87,0.9642)(88,0.9643)(89,0.9643)(90,0.9645)(91,0.9645)(92,0.9646)(93,0.9647)(94,0.9648)(95,0.9648)(96,0.9649)(97,0.965)(98,0.965)(99,0.965)(100,0.9651)};

        \nextgroupplot[xlabel={SSRL Training Epoch}, ylabel={Accuracy (\%)}, xtick={0, 50, 100}, xticklabels={0, 50, 100}, ytick={56.95, 66.07, 78.16, 88.08}, yticklabels={56.95, 66.07, 78.16, 88.08}]
            \addplot+[myBlue, line width=0.6pt, mark=none] coordinates {(0, 56.95) (1, 58.81) (2, 59.63) (3, 61.83) (4, 63.79) (5, 65.0) (6, 65.99) (7, 66.94) (8, 67.81) (9, 68.53) (10, 69.16) (11, 69.72) (12, 70.22) (13, 70.69) (14, 71.13) (15, 71.54) (16, 71.91) (17, 72.23) (18, 72.54) (19, 72.84) (20, 73.11) (21, 73.36) (22, 73.6) (23, 73.83) (24, 74.03) (25, 74.23) (26, 74.4) (27, 74.57) (28, 74.74) (29, 74.9) (30, 75.05) (31, 75.2) (32, 75.33) (33, 75.45) (34, 75.57) (35, 75.67) (36, 75.79) (37, 75.89) (38, 75.98) (39, 76.08) (40, 76.17) (41, 76.26) (42, 76.34) (43, 76.42) (44, 76.5) (45, 76.57) (46, 76.63) (47, 76.69) (48, 76.76) (49, 76.83) (50, 76.89) (51, 76.96) (52, 77.02) (53, 77.08) (54, 77.14) (55, 77.19) (56, 77.24) (57, 77.29) (58, 77.33) (59, 77.38) (60, 77.42) (61, 77.46) (62, 77.51) (63, 77.55) (64, 77.6) (65, 77.63) (66, 77.66) (67, 77.69) (68, 77.73) (69, 77.75) (70, 77.77) (71, 77.8) (72, 77.82) (73, 77.85) (74, 77.86) (75, 77.88) (76, 77.9) (77, 77.92) (78, 77.94) (79, 77.95) (80, 77.97) (81, 77.99) (82, 78.01) (83, 78.01) (84, 78.03) (85, 78.04) (86, 78.06) (87, 78.06) (88, 78.07) (89, 78.08) (90, 78.09) (91, 78.09) (92, 78.1) (93, 78.11) (94, 78.12) (95, 78.12) (96, 78.13) (97, 78.14) (98, 78.15) (99, 78.15) (100, 78.16)};
          \addplot+[myRed, line width=0.6pt, mark=none] coordinates{(0,66.07)(1,66.49)(2,67.01)(3,68.96)(4,71.85)(5,73.68)(6,75.08)(7,76.31)(8,77.36)(9,78.26)(10,79.03)(11,79.71)(12,80.33)(13,80.9)(14,81.4)(15,81.87)(16,82.31)(17,82.71)(18,83.07)(19,83.39)(20,83.7)(21,83.98)(22,84.22)(23,84.43)(24,84.62)(25,84.77)(26,84.92)(27,85.06)(28,85.17)(29,85.27)(30,85.38)(31,85.48)(32,85.56)(33,85.63)(34,85.69)(35,85.74)(36,85.79)(37,85.84)(38,85.89)(39,85.92)(40,85.96)(41,85.99)(42,86.03)(43,86.06)(44,86.09)(45,86.11)(46,86.12)(47,86.14)(48,86.15)(49,86.16)(50,86.19)(51,86.23)(52,86.29)(53,86.41)(54,86.53)(55,86.65)(56,86.79)(57,86.92)(58,87.05)(59,87.17)(60,87.28)(61,87.37)(62,87.46)(63,87.53)(64,87.59)(65,87.64)(66,87.69)(67,87.73)(68,87.77)(69,87.8)(70,87.83)(71,87.85)(72,87.87)(73,87.89)(74,87.9)(75,87.92)(76,87.93)(77,87.95)(78,87.96)(79,87.97)(80,87.98)(81,87.99)(82,88.0)(83,88.0)(84,88.01)(85,88.02)(86,88.03)(87,88.03)(88,88.04)(89,88.04)(90,88.05)(91,88.06)(92,88.06)(93,88.06)(94,88.06)(95,88.06)(96,88.07)(97,88.07)(98,88.07)(99,88.08)(100,88.08)};

        \nextgroupplot[xlabel={SSRL Training Epoch}, ylabel={Purity (\%)}, xtick={0, 50, 100}, xticklabels={0, 50, 100}, ytick={65.58, 87.47, 92.1}, yticklabels={65.58, 87.47, 92.10}]
            \addplot+[myBlue, line width=0.6pt, mark=none] coordinates {(0, 65.58) (1, 68.88) (2, 69.11) (3, 69.51) (4, 72.04) (5, 75.73) (6, 77.52) (7, 78.59) (8, 79.62) (9, 80.42) (10, 81.07) (11, 81.64) (12, 82.04) (13, 82.46) (14, 82.88) (15, 83.17) (16, 83.47) (17, 83.64) (18, 83.84) (19, 84.03) (20, 84.19) (21, 84.37) (22, 84.51) (23, 84.57) (24, 84.7) (25, 84.86) (26, 84.99) (27, 85.08) (28, 85.22) (29, 85.29) (30, 85.4) (31, 85.48) (32, 85.57) (33, 85.64) (34, 85.7) (35, 85.79) (36, 85.88) (37, 85.94) (38, 86.01) (39, 86.07) (40, 86.11) (41, 86.15) (42, 86.22) (43, 86.25) (44, 86.31) (45, 86.36) (46, 86.41) (47, 86.48) (48, 86.53) (49, 86.58) (50, 86.63) (51, 86.67) (52, 86.71) (53, 86.75) (54, 86.79) (55, 86.83) (56, 86.85) (57, 86.88) (58, 86.9) (59, 86.92) (60, 86.94) (61, 86.97) (62, 86.99) (63, 87.03) (64, 87.05) (65, 87.07) (66, 87.08) (67, 87.11) (68, 87.14) (69, 87.15) (70, 87.16) (71, 87.17) (72, 87.18) (73, 87.2) (74, 87.22) (75, 87.24) (76, 87.26) (77, 87.27) (78, 87.29) (79, 87.3) (80, 87.31) (81, 87.32) (82, 87.33) (83, 87.33) (84, 87.35) (85, 87.35) (86, 87.36) (87, 87.38) (88, 87.38) (89, 87.39) (90, 87.39) (91, 87.39) (92, 87.39) (93, 87.42) (94, 87.42) (95, 87.43) (96, 87.44) (97, 87.46) (98, 87.47) (99, 87.48) (100, 87.47)};
          \addplot+[myRed, line width=0.6pt, mark=none] coordinates{(0,88.75)(1,88.81)(2,88.87)(3,88.79)(4,88.89)(5,89.43)(6,90.13)(7,90.38)(8,90.59)(9,90.79)(10,90.82)(11,90.9)(12,90.96)(13,90.99)(14,90.97)(15,90.95)(16,90.96)(17,90.98)(18,90.95)(19,90.94)(20,91.01)(21,91.04)(22,91.04)(23,91.08)(24,91.12)(25,91.17)(26,91.19)(27,91.22)(28,91.26)(29,91.3)(30,91.35)(31,91.4)(32,91.43)(33,91.48)(34,91.51)(35,91.53)(36,91.56)(37,91.58)(38,91.6)(39,91.62)(40,91.65)(41,91.66)(42,91.68)(43,91.7)(44,91.73)(45,91.76)(46,91.8)(47,91.84)(48,91.86)(49,91.86)(50,91.84)(51,91.83)(52,91.78)(53,91.74)(54,91.68)(55,91.63)(56,91.59)(57,91.54)(58,91.51)(59,91.48)(60,91.47)(61,91.46)(62,91.43)(63,91.44)(64,91.42)(65,91.44)(66,91.47)(67,91.49)(68,91.51)(69,91.52)(70,91.55)(71,91.57)(72,91.6)(73,91.63)(74,91.66)(75,91.67)(76,91.69)(77,91.72)(78,91.73)(79,91.77)(80,91.79)(81,91.82)(82,91.85)(83,91.87)(84,91.89)(85,91.91)(86,91.91)(87,91.92)(88,91.94)(89,91.97)(90,91.99)(91,91.99)(92,92.01)(93,92.03)(94,92.04)(95,92.05)(96,92.07)(97,92.09)(98,92.08)(99,92.09)(100,92.1)};
    \end{groupplot}
    \node (legend) at ($(group c4r1.north)+(-2.15cm,0.1cm)$) [above] {\ref{grouplegend}};
\end{tikzpicture}
    \caption{Evolution of pseudo labeling across training epochs during the SSRL training phase.}
    \label{fig:labelingvsepoch}
\end{figure*}

\subsubsection{Efficiency of the SSRL approach}
Unlike the iterative second stage which requires multiple training rounds, SSRL introduces a more streamlined approach. This eliminates the need for iterative training, leading to improved efficiency.
This is illustrated in Figure~\ref{fig:eervstime}, which compares the EER over training time between the iterative second stage and SSRL second stage in the two-stage framework.\footnote{All models are trained on two NVIDIA GeForce RTX 3090 GPUs. The estimated training time focuses solely on ideal conditions, accounting only for the forward and backward propagation time (model training time of a single batch). It excludes time allocations for data loading, preprocessing pipeline, model validations, and procedures like k-means clustering and GMM modeling. These processes, being brief in nature, are considered negligible.}

For the ResNet-based pipeline, it is apparent from the visualization that SSRL achieves quicker convergence and maintains a more stable EER than iterative training. The iterative approach exhibits fluctuations due to its clustering process, where pseudo labels are re-generated between rounds, requiring random initialization of the final linear layer. This causes temporary EER spikes before stabilization. In contrast, SSRL continuously refines pseudo labels within a single training round, enabling smoother training dynamics and improved efficiency. These advantages demonstrate SSRL's potential for applications where training time and computational resources are critical considerations.

The ECAPA-TDNN-based pipeline fails to converge and experiences overfitting during each training round in the iterative second stage.
The verification performance (EER) shows minimal improvement before rapidly deteriorating.
This occurs because we initialize the network using parameters from the previous round and k-means centers for the final linear layer, causing rapid data fitting in early epochs.
Due to this overfitting tendency, we terminated training after the third round.
These results suggest that with a strong initialization (DINO pretrained model), the two-stage iterative framework cannot substantially enhance performance.
In contrast, the proposed SSRL method dynamically adjusts clustering, leading to further performance improvements even when starting with a relatively well-pretrained model.

\begin{table}[t!]
  \caption{Comparison of the proposed SSRL method with two-stage iterative framework variants. EERs [\%] from VoxCeleb 1-O test trial; all models use ECAPA-TDNN. `Filter' denotes mislabeled sample filtering, `LC' for label correction.}
  \label{tab: ssrl result3}
  \centering
  \begin{tabular}[c]{@{}l@{\ }c@{\ }c@{\ }c@{}c@{}c@{\ }c@{\ }c@{}}
    \toprule
    Method & Loss & Filter & LC & Other & \#Rounds & \tabincell{c}{Stage 1\\EER} & EER \\
    \midrule
    Thienpondt \textit{et al.} \cite{thienpondtidlab} & AAM & - & - & - & 7 & 7.3 & 2.1 \\
    Mun \textit{et al.} \cite{mun2021snu} & AAM & - & - & score norm & 5 & 3.65 & 1.66 \\
    Tao \textit{et al.} \cite{tao_self-supervised_2022} & AAM & \checkmark & - & - & 5 & 7.36 & 1.66 \\
    Han \textit{et al.} \cite{han_self-supervised_2022} & AAM & \checkmark & \checkmark & - & 5 & 6.16 & 1.47 \\
    Tao \textit{et al.} \cite{tao2023self} & AAM & - & - & audio-visual & $\geq$2 & \textbf{2.89} & 1.44 \\
    Chen \textit{et al.} \cite{chen_self-supervised_2023} & AAM & - & - & audio-visual & 7 & 7.16 & 1.27 \\
    Chen \textit{et al.} \cite{chen_unsupervised_2023} & AAM & - & \checkmark & WavLM & 5 & - & \textbf{1.25} \\
    \midrule
    SSRL (proposed) & CE & - & \checkmark & - & 1 & 2.94 & 1.77 \\
    SSRL (proposed) & AAM & - & \checkmark & - & 1 & 2.94 & \textbf{1.25} \\
    SSRL (proposed) & AAM & - & \checkmark & WavLM & 1 & 2.94 & \textbf{1.04} \\
    \bottomrule
  \end{tabular}
\end{table}

\subsubsection{Comparative analysis with other two-stage iterative framework variants}

In Table \ref{tab: ssrl result3}, the performance of the proposed SSRL method is compared with various two-stage iterative framework variants, all leveraging the ECAPA-TDNN model.
A remarkable observation is the efficiency and efficacy of SSRL when trained with the AAM loss: it surpasses all other methods, achieving superior performance within a single training round.

Two comparisons deserve special mention.
First, when contrasted with the work of Chen \textit{et al.} \cite{chen_self-supervised_2023} -- which incorporates an additional visual modality during training -- our SSRL method delivers performance on par, even though it relies exclusively on audio information.
Secondly, another variant from Chen \textit{et al.} \cite{chen_unsupervised_2023} makes use of a subset of WavLM \cite{9814838}, a large self-supervised speech model trained on extensive data, for feature extraction.
Our SSRL approach, devoid of any large-scale pre-trained model, emerges with a similar performance.

To further evaluate our approach, we integrated WavLM as a feature extractor alongside the proposed SSRL method. Specifically, we extracted features from every layers of WavLM-Large encoder and combined them using a learnable weighted sum to create composite features for input to ECAPA-TDNN. Our training strategy involved initially freezing the WavLM parameters during early epochs, followed by gradual fine-tuning. This integration proved highly effective: the system achieved 1.04\% EER on Vox1-O, representing a significant 16.8\% improvement over the SSRL model trained with Mel-filterbank features (1.25\% EER).

\subsection{Pseudo labeling performance}
\label{sec: labeling}
Table \ref{tab: ssrl labeling} details the pseudo labeling performance of the ResNet and ECAPA models on the training data.
The SSRL method consistently shows superior metrics across both model architectures.
For instance, with the ResNet model, the SSRL technique achieved an accuracy of 78.12\% compared to the 68.93\% from the two-stage iterative framework in its fifth training round. 
This superior performance can be attributed to the online clustering achieved through self-supervised knowledge distillation, coupled with additional strategies to enhance the quality of pseudo labels.

\begin{table}[t!]
  \caption{Pseudo labeling performance on training data.}
  \label{tab: ssrl labeling}
  \centering
  \begin{tabular}[c]{@{\ }l@{\ \ }l@{\ \ }c@{\ \ }c@{\ \ }c@{\ \ }c@{\ }}
    \toprule
    Model & Method & \#Clusters & NMI & Accuracy & Purity \\
    \midrule
    \multirow{2}*{ResNet}
    & Iterative round 5 & 6000 & 0.9230 & 68.93\% & 83.50\% \\
    & SSRL & 8000$\rightarrow$5085 & 0.9333 & 78.12\% & 87.42\% \\
    \midrule
    \multirow{2}*{ECAPA} 
    & Iterative round 3 & 8000 & 0.9333 & 64.83\% & 89.35\% \\
    & SSRL & 8000$\rightarrow$5328 & 0.9651 & 88.08\% & 92.10\%\\ 
    \bottomrule
  \end{tabular}
\end{table}

\begin{table*}[t!]
  \caption{Performance comparison of the SSRL approach with different component configurations. $p_\text{clean}$ represents the proposed noisy label modeling method. $K$ represents the converged number of clusters. The actual cluster counts is 5994.}
  \label{tab: ssrl ab1}
  \centering
  \begin{tabular}[c]{cccc ccc cccc}
    \toprule
    \multirow{2}*{\tabincell{@{}c@{}}{Online\\Clustering}} & \multirow{2}*{EMA} & \multirow{2}*{\tabincell{@{}c@{}}{Label\\Queue}} & \multirow{2}*{$p_\text{clean}$} & \multicolumn{3}{@{}c@{}}{Verification EER[\%] \textdownarrow} & \multicolumn{4}{@{}c@{}}{Pseudo Labeling} \\ 
    \cmidrule(lr){5-7} \cmidrule(lr){8-11}
    & & & & Vox1-O & Vox1-E & Vox1-H & NMI \textuparrow & Acc \textuparrow & Purity \textuparrow & $K$ \\
    \midrule
    argmax & \checkmark & \checkmark & \checkmark & \textbf{2.39} & 2.63 & 4.74 & 0.9333 & 78.12\% & 87.42\% & 5085 \\
    argmax & \ding{55} & \checkmark & \checkmark & 2.48 & 2.97 & 5.31 & 0.9261 & 77.23\% & 87.76\% & 4943 \\
    argmax & \checkmark & \ding{55} & \checkmark & 2.51 & 2.68 & 4.82 & 0.9297 & 77.31\% & 86.96\% & 4801 \\
    argmax & \checkmark & \checkmark & \ding{55} & 2.76 & 2.94 & 5.29 & 0.9300 & 75.55\% & 86.55\% & 6152 \\
    argmax & \ding{55} & \ding{55} & \ding{55} & 5.19 & 6.52 & 11.73 & 0.8567 & 66.88\% & \textbf{90.64\%} & 4306 \\    
    Sinkhorn & \checkmark & \checkmark & \checkmark & 2.41 & \textbf{2.57} & \textbf{4.61} & \textbf{0.9402} & \textbf{79.26\%} & 84.45\% & 5974 \\
    \bottomrule
  \end{tabular}
\end{table*}

In Figure \ref{fig:labelingvsepoch}, the evolution of pseudo labeling throughout the training epochs using the SSRL method is depicted.
As observed, during the SSRL training process, there's a consistent reduction in the number of clusters across epochs until a stable count is reached.
For the ResNet-based SSRL, this stable number is 5084, whereas for the ECAPA-TDNN-based SSRL, it's 5328.
For reference, the training data contains a total of 5994 speakers.
These observations indicate that the SSRL method is adept at filtering out pseudo clusters that have lower confidence, thereby progressively optimizing the labeling performance.
Moreover, metrics such as NMI, accuracy, and maximal purity per cluster show that with each passing epoch, the SSRL-trained models fine-tune their performance, reflecting continuous improvement.

\subsection{Ablation study}

The SSRL approach employs components designed to enhance the model's performance on the unlabeled dataset.
To inspect the contributions of these components, we conduct ablation studies on the ResNet-based SSRL, shown in Table \ref{tab: ssrl ab1}.

\subsubsection{Speaker verification performance analysis}

When the EMA update for the teacher model is integrated into SSRL, we observe an improvement in speaker verification performance, with EERs of 2.39\%, 2.63\%, and 4.74\% across the VoxCeleb test trials. In contrast, the model without the EMA update shows higher EERs of 2.48\%, 2.97\%, and 5.31\%, respectively. This empirical evidence underscores the crucial role of EMA in SSRL for enhancing speaker verification performance.

Furthermore, the pseudo label queue further improves the SSRL model. 
Its integration not only amplifies speaker verification capabilities but also buffers against potential pitfalls associated with pseudo labeling. From Table \ref{tab: ssrl ab1}, we can see that the experiment without using the label queue results in worse EERs and pseudo-labeling performance compared to the one that includes it. Specifically, the final cluster count (4,801) is significantly smaller, indicating that many classes were removed during training. This suggests that training with noisy labels leads to bias toward certain classes, and without correction, the model reinforces this bias, causing pseudo labels to collapse into fewer clusters. The label queue serves as a buffer against these potential pitfalls by stabilizing pseudo labels and preventing the excessive merging of speaker identities.

Notably, introduction of noisy label modeling with $p_\text{clean}$ provides an additional layer of refinement to the SSRL approach.
By guiding predictions towards cleaner samples, this mechanism mitigates the challenges associated with noisy label updates.
A degradation in speaker verification performance is observed in the absence of noisy label modeling, with EER increase to 2.76\%, 2.94\%, and 5.29\% across the test trials.

\begin{figure}
    \centering
    \includegraphics[width=\linewidth]{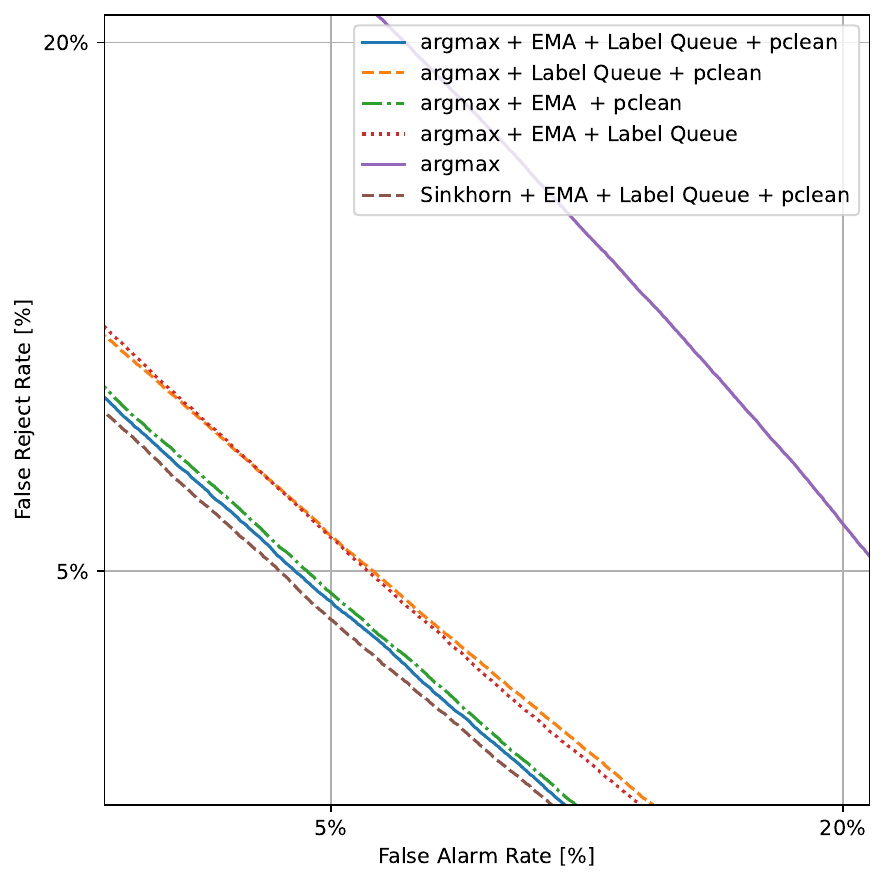}
    \caption{DET curves on VoxCeleb 1-H test trial: comparing SSRL methods with different component configurations.}
    \label{fig: det plot}
\end{figure}

Additionally, we evaluated a simplified version of SSRL.
This variant, devoid of the EMA updates, pseudo label queue, and noisy label modeling, preserves only the noisy student training strategy.
As observed in Table VI, this simplified version undergoes a significant performance drop, with an EER increase of 2.8 percentage points (2.39\% $\rightarrow$ 5.19\%) compared to the full SSRL approach on the VoxCeleb 1-O test trial. In fact, this simplified SSRL had difficulty converging.
These observations underscore the collective significance of the various components in achieving optimal performance with SSRL. The detection error tradeoff (DET) plots on VoxCeleb 1-H test trial is shown in Figure \ref{fig: det plot} to compare the SSRL approach with different component configurations.

\begin{table*}[th!]
  \caption{Performance comparison of the SSRL approach using varying numbers of clusters for the initial pseudo labels. $K_\mathrm{init}$ denotes the initial number of clusters; $K_\mathrm{converged}$ indicates the number of clusters upon convergence. The arrow illustrates the transition from the model trained with the fixed initial clustering for 50 epochs to the converged SSRL.}
  \label{tab: ssrl ab3}
  \centering
  \begin{tabular}[c]{l ccc cccc}
    \toprule
    \multirow{2}*{$K_\text{init}$} & \multicolumn{3}{@{}c@{}}{Verification EER[\%] \textdownarrow} & \multicolumn{4}{@{}c@{}}{Pseudo Labeling} \\ 
    \cmidrule(lr){2-4} \cmidrule(lr){5-8}
    & Vox1-O & Vox1-E & Vox1-H & NMI \textuparrow & Acc \textuparrow & Purity \textuparrow & $K_\text{converged}$\\
    \midrule
    1,000\footnotemark & \rotatebox{90}{$\curvearrowleft$} \tabincell{@{}c@{}}{5.98\\6.73} & \rotatebox{90}{$\curvearrowleft$} \tabincell{@{}c@{}}{6.51\\7.70} & \rotatebox{90}{$\curvearrowleft$} \tabincell{@{}c@{}}{11.87\\13.93} & \rotatebox{90}{$\curvearrowleft$} \tabincell{@{}c@{}}{0.6189\\0.6658} & \rotatebox{90}{$\curvearrowleft$} \tabincell{@{}c@{}}{19.84\%\\21.84\%} & \rotatebox{90}{$\curvearrowleft$} \tabincell{@{}c@{}}{21.58\%\\45.58\%} & 714 \\
    \midrule
    8,000 & \rotatebox{90}{$\curvearrowleft$} \tabincell{@{}c@{}}{4.05\\\textbf{2.39}} & \rotatebox{90}{$\curvearrowleft$} \tabincell{@{}c@{}}{4.61\\\textbf{2.63}} & \rotatebox{90}{$\curvearrowleft$} \tabincell{@{}c@{}}{8.58\\\textbf{4.74}} & \rotatebox{90}{$\curvearrowleft$} \tabincell{@{}c@{}}{0.7744\\0.9333} & \rotatebox{90}{$\curvearrowleft$} \tabincell{@{}c@{}}{36.87\%\\\textbf{78.12\%}} & \rotatebox{90}{$\curvearrowleft$} \tabincell{@{}c@{}}{55.32\%\\87.42\%} & 5,085 \\
    \midrule
    20,000 & \rotatebox{90}{$\curvearrowleft$} \tabincell{@{}c@{}}{3.94\\3.03} & \rotatebox{90}{$\curvearrowleft$} \tabincell{@{}c@{}}{4.29\\2.92} & \rotatebox{90}{$\curvearrowleft$} \tabincell{@{}c@{}}{7.88\\5.21} & \rotatebox{90}{$\curvearrowleft$} \tabincell{@{}c@{}}{0.8114\\\textbf{0.9356}} & \rotatebox{90}{$\curvearrowleft$} \tabincell{@{}c@{}}{27.68\%\\73.35\%} & \rotatebox{90}{$\curvearrowleft$} \tabincell{@{}c@{}}{68.21\%\\\textbf{92.66\%}} & 9,956 \\
    \bottomrule
  \end{tabular}
\end{table*}

\subsubsection{Pseudo labeling analysis}
The dynamism inherent in the online clustering mechanism deserves mention.
Through SSRL's online clustering, certain clusters are filtered out or merged as training progresses, thereby stabilizing the number of clusters towards the conclusion.
Referring to Table \ref{tab: ssrl ab1}, the full version of SSRL, equipped with the EMA update, pseudo label queue, and noisy label modeling, has a converged cluster count of 5085.
However, models without either the EMA update or the pseudo label queue end with smaller cluster counts, registering at 4943 and 4801, respectively.
This observation indicates that the EMA update and pseudo label queue jointly act as regularizers for pseudo cluster prediction, fostering stability throughout the training epochs, thus preventing the cluster counts shrink too quickly.
Specifically, the pseudo label queue serves as a buffer against erratic predictions, enhancing stability and minimizing outliers, while the EMA component ensures that the network remains consistent in its cluster assignment predictions.

Conversely, the noisy label modeling approach appears to have an opposing effect on the converged cluster count compared to the EMA update and pseudo label queue.
Excluding noisy label modeling culminates in an increased cluster count of 6152.
This suggests that noisy label modeling prioritizes predictions for more confident clusters, diminishing those with less confidence, which consequently reduces the overall cluster count.

\footnotetext{When trained with an initial cluster count of 1,000, the model could not converge, so we stopped the training after 50 epochs of SSRL.}

\begin{table}[t!]
  \caption{Performance comparison of the SSRL approach with different pseudo label queue length $L$. $K$ represents the converged number of clusters.}
  \label{tab: ssrl ab2}
  \centering
  \begin{tabular}[c]{@{\ }l c@{\ \ \ }c@{\ \ \ }c c@{\ \ \ }c@{\ \ \ }c@{\ \ \ }c@{\ }}
    \toprule
    \multirow{2}*{$L$} & \multicolumn{3}{@{}c@{}}{Verification EER[\%] \textdownarrow} & \multicolumn{4}{@{}c@{}}{Pseudo Labeling} \\ 
    \cmidrule(lr){2-4} \cmidrule(lr){5-8}
    & Vox1-O & Vox1-E & Vox1-H & NMI \textuparrow & Acc \textuparrow  & Purity \textuparrow & $K$\\
    \midrule
    1\tablefootnote{Label queue method is disabled when label queue length $L=1$.} & 2.51 & 2.68 & 4.82 & 0.9297 & 77.31\% & 86.96\% & 4801 \\
    5 & \textbf{2.39} & \textbf{2.63} & \textbf{4.74} & \textbf{0.9333} & \textbf{78.12\%} & \textbf{87.42\%} & 5085 \\
    10 & 2.45 & 2.65 & 4.79 & 0.9322 & 77.58\% & 87.31\% & 5320 \\
    20 & 2.56 & 2.73 & 4.93 & 0.9296 & 76.87\% & 86.97\% & 5485 \\
    \bottomrule
  \end{tabular}
\end{table}

\subsubsection{Direct maximum probability versus Sinkhorn-based online clustering}
To investigate the impact of different online clustering techniques, we evaluated two approaches: direct maximum probability assignment and cluster assignment through optimal transport.

In terms of speaker verification, both methods prove efficacious, achieving comparable EERs across test trials. On VoxCeleb 1-E and VoxCeleb 1-H, the Sinkhorn approach registers marginal improvements with EERs of 2.57\% and 4.61\% compared to 2.63\% and 4.74\% using direct assignment. This suggests that the two techniques are largely comparable in enhancing speaker verification capabilities.

Regarding pseudo labeling, the Sinkhorn method manifests an edge, garnering superior metrics of clustering accuracy (79.26\% vs 78.12\%) and NMI (0.9402 vs 0.9333). This indicates an enhanced capacity for accurate pseudo label generation using the optimal transport approach.
Inspecting the converged number of clusters, Sinkhorn retains more clusters upon convergence at 5974, contrasted with 5085 using direct assignment.
This aligns with the constraint in the Sinkhorn algorithm to distribute samples evenly across clusters.
Conversely, the direct assignment aggressively merges smaller, outlier clusters.
In summary, both online clustering techniques prove effective and validate the online clustering mechanism's efficacy in SSRL.

\subsubsection{Interplay of initial cluster count}
Table \ref{tab: ssrl ab3} shows the interplay between the initial cluster count $K_\text{init}$ and the SSRL approach's performance.
An overly conservative choice for $K_\text{init}$ (e.g., 1,000) seems to restrict the model's ability to capture the data's inherent diversity, leading to suboptimal results.
In contrast, an overly aggressive $K_\text{init}$ (e.g., 20,000) does allow for improved pseudo labeling metrics, but doesn't necessarily translate to the best verification EER.
In summary, the choice of $K_\text{init}$ is crucial.
It acts as a balance between providing enough granularity for capturing data diversity and ensuring the model remains focused on meaningful clusters.

\subsubsection{Impact of pseudo label queue length $L$}
Table \ref{tab: ssrl ab2} shows the impact of pseudo label queue length $L$ on the model's performance. 
The pseudo label queue filters transient inconsistencies, and ensuring continuity in predicted pseudo labels across training epochs.
An observation is the marginal degradation in performance as $L$ increases beyond a certain threshold.
With $L=1$, essentially indicating no pseudo label queue, the verification EER on VoxCeleb 1-O test trial is 2.51\% and the converged number of clusters $K$ stands at 4801.
Increasing $L$ to 5 yields a better EER of 2.39\% and a higher $K$ of 5084.
Further increments in $L$ to 10 and 20, however, show worse EERs and expanding $K$s.
This trend suggests an optimal range for $L$ where the benefits of temporal stabilization maximize.
An excessively long queue might integrate older, potentially less relevant pseudo labels, causing slight deteriorations in performance.
This observation aligns with the inherent trade-off: while having some history aids in stabilization, overly long histories might dilute the recent advancements the model has achieved.

\subsection{Fine-tuning}
In this section, the SSRL pre-trained ECAPA-TDNN speaker model is fine-tuned with small-scale labeled datasets. We use the VoxCeleb 1 development set (1,211 speakers) \cite{nagrani_voxceleb:_2017} for fine-tuning and create an additional subset of 600 randomly selected speakers to evaluate self-supervised pre-training on smaller datasets. Results are reported on the VoxCeleb 1-O test trials.

As shown in Table \ref{tab:ft}, fine-tuning the SSRL model with labeled data significantly improves performance: fine-tuning on only 600 speakers achieves an EER of 1.05\%, compared to 3.94\% without SSRL pre-training. Fine-tuning the SSRL model on all labeled speakers in VoxCeleb 1 (1,211 speakers) further reduces the EER to 0.95\%, compared to 2.31\% without SSRL pre-training. These results demonstrate that SSRL provides a strong self-supervised foundation, which can be further enhanced with labeled data for improved speaker verification.

\begin{table}[t]
  \caption{Fine-tune the self-supervised model with different labeled data in VoxCeleb 1 development set.}
  \label{tab:ft}
  \centering
  \begin{tabular}[c]{@{}l@{\ }c@{\ }c@{\ }c@{\ }c@{\ }c@{\ }c@{}}
    \toprule
    \textbf{Fine-tuning Data}  & \multicolumn{2}{@{}c@{}}{None} & \multicolumn{2}{@{}c@{}}{600 Speakers} & \multicolumn{2}{@{}c@{}}{1,211 Speakers} \\ 
    \cmidrule(lr){2-3} \cmidrule(lr){4-5} \cmidrule(lr){6-7}
    \textbf{Pre-trained Model} & minDCF & EER[\%] & minDCF & EER[\%] & minDCF & EER[\%] \\
    \midrule
    None & - & - & 0.295 & 3.94 & 0.175 & 2.31\\ 
    SSRL & 0.101 & 1.25 & 0.089 & 1.05 & 0.075 & 0.95 \\ 
    \bottomrule
  \end{tabular}
\end{table}

\section{Conclusion}

This paper introduces self-supervised reflective learning (SSRL), a novel paradigm for unsupervised speaker representation learning. SSRL streamlines existing two-stage iterative frameworks by integrating self-supervised knowledge distillation with online clustering. A teacher model continually refines pseudo labels through clustering, providing dynamic supervision to train the student model. The method also employs techniques like label correction and noisy label modeling to further improve pseudo label quality.

Our experiments demonstrate SSRL's superiority over current two-stage iterative approaches. On VoxCeleb 1 test trials, SSRL surpasses the performance of a 5-round iterative method in just a single training round. Ablation studies validate the contributions of key components like noisy label modeling, pseudo label queues, and EMA teacher updates.
Moreover, the consistent improvement in pseudo labeling throughout the training phase, coupled with the convergence of cluster count, reaffirms SSRL's prowess in deciphering pertinent clusters within unlabeled data.

This work marks a pivotal advancement in efficient and accurate speaker representation learning. By combining self-supervised distillation and online clustering, SSRL eliminates previous iterative bottlenecks. The reflective learning paradigm introduces new horizons for developing scalable, unsupervised systems. Future work should assess SSRL on larger datasets and expand hyperparameter optimizations. Integrating SSRL into end-to-end pipelines is another research direction.

\section{Acknowledgments}
This research is funded in part by the National Natural Science Foundation of China (62171207), Science and Technology Program of Suzhou City (SYC2022051) and Guangdong Science and Technology Plan (2023A1111120012). Many thanks for the computational resource provided by the Advanced Computing East China Sub-Center.

\bibliographystyle{IEEEtran}
\bibliography{mybib.bib}

\begin{thebibliography}{10}
\providecommand{\url}[1]{#1}
\csname url@samestyle\endcsname
\providecommand{\newblock}{\relax}
\providecommand{\bibinfo}[2]{#2}
\providecommand{\BIBentrySTDinterwordspacing}{\spaceskip=0pt\relax}
\providecommand{\BIBentryALTinterwordstretchfactor}{4}
\providecommand{\BIBentryALTinterwordspacing}{\spaceskip=\fontdimen2\font plus
\BIBentryALTinterwordstretchfactor\fontdimen3\font minus \fontdimen4\font\relax}
\providecommand{\BIBforeignlanguage}[2]{{%
\expandafter\ifx\csname l@#1\endcsname\relax
\typeout{** WARNING: IEEEtran.bst: No hyphenation pattern has been}%
\typeout{** loaded for the language `#1'. Using the pattern for}%
\typeout{** the default language instead.}%
\else
\language=\csname l@#1\endcsname
\fi
#2}}
\providecommand{\BIBdecl}{\relax}
\BIBdecl

\bibitem{chen_large-scale_2022}
Z.~Chen, S.~Chen, Y.~Wu, Y.~Qian, C.~Wang, S.~Liu, Y.~Qian, and M.~Zeng, ``{Large-Scale} {{Self-Supervised Speech Representation Learning}} for {{Automatic Speaker Verification}},'' in \emph{{{Proceeding of ICASSP}}}, 2022, pp. 6147--6151.

\bibitem{chen_pushing_2023}
Y.~Chen, S.~Zheng, H.~Wang, L.~Cheng, and Q.~Chen, ``Pushing the {{Limits}} of {{Self-Supervised Speaker Verification Using Regularized Distillation Framework}},'' in \emph{{{Proceeding of ICASSP}}}, 2023, pp. 1--5.

\bibitem{tu2024contrastive}
Y.~Tu, M.-W. Mak, and J.-T. Chien, ``{Contrastive Self-Supervised Speaker Embedding with Sequential Disentanglement},'' \emph{IEEE/ACM Transactions on Audio, Speech, and Language Processing}, 2024.

\bibitem{liu2023self}
Y.~Liu, L.-F. Wei, C.-F. Zhang, T.-H. Zhang, S.-L. Chen, and X.-C. Yin, ``{Self-Supervised Contrastive Speaker Verification with Nearest Neighbor Positive Instances},'' \emph{Pattern Recognition Letters}, vol. 173, pp. 17--22, 2023.

\bibitem{10448455}
Z.~Zhou, H.~Yang, and T.~Shinozaki, ``{Self-Supervised Speaker Verification with Adaptive Threshold and Hierarchical Training},'' in \emph{Proceeding of ICASSP}, 2024, pp. 12\,141--12\,145.

\bibitem{fathan2024analytic}
A.~Fathan and J.~Alam, ``{An Analytic Study on Clustering Driven Self-Supervised Speaker Verification},'' \emph{Pattern Recognition Letters}, vol. 179, pp. 80--86, 2024.

\bibitem{wang2024leveraging}
S.~Wang, Q.~Bai, Q.~Liu, J.~Yu, Z.~Chen, B.~Han, Y.~Qian, and H.~Li, ``Leveraging in-the-wild data for effective self-supervised pretraining in speaker recognition,'' in \emph{Proceeding of ICASSP}, 2024, pp. 10\,901--10\,905.

\bibitem{cai_iterative_2021}
D.~Cai, W.~Wang, and M.~Li, ``An {{Iterative Framework}} for {{Self}}-{{Supervised Deep Speaker Representation Learning}},'' in \emph{Proceeding of ICASSP}, 2021, pp. 6728--6732.

\bibitem{9741340}
D.~\vspace{0mm}Cai, W.~Wang, and M.~Li, ``{Incorporating Visual Information in Audio Based Self-Supervised Speaker Recognition},'' \emph{IEEE/ACM Transactions on Audio, Speech, and Language Processing}, vol.~30, pp. 1422--1435, 2022.

\bibitem{chen_exploring_2021}
X.~Chen and K.~He, ``Exploring {{Simple Siamese Representation Learning}},'' in \emph{{{Proceedings of CVPR}}}, 2021, pp. 15\,750--15\,758.

\bibitem{hinton_distilling_2015}
G.~Hinton, O.~Vinyals, and J.~Dean, ``Distilling the {{Knowledge}} in a {{Neural Network}},'' in \emph{{{NeurIPS Deep Learning}} and {{Representation Learning Workshop}}}, 2015.

\bibitem{grill_bootstrap_2020}
J.-B. Grill, F.~Strub, F.~Altch{\'e}, C.~Tallec, P.~Richemond, E.~Buchatskaya, C.~Doersch, B.~Avila~Pires, Z.~Guo, M.~Gheshlaghi~Azar \emph{et~al.}, ``Bootstrap {{Your Own Latent}}: {{A New Approach}} to {{Self-Supervised Learning}},'' \emph{NeurIPS}, vol.~33, pp. 21\,271--21\,284, 2020.

\bibitem{caron_emerging_2021}
M.~Caron, H.~Touvron, I.~Misra, H.~J{\'e}gou, J.~Mairal, P.~Bojanowski, and A.~Joulin, ``{Emerging Properties in Self-supervised Vision Transformers},'' in \emph{{{Proceedings of ICCV}}}, 2021, pp. 9650--9660.

\bibitem{arazo_unsupervised_2019}
E.~Arazo, D.~Ortego, P.~Albert, N.~E. O'Connor, and K.~McGuinness, ``Unsupervised {{Label Noise Modeling}} and {{Loss Correction}},'' in \emph{{{Proceedings of the International Conference on Machine Learning}}}, 2019.

\bibitem{elbanna2022byol}
G.~Elbanna, N.~Scheidwasser-Clow, M.~Kegler, P.~Beckmann, K.~El~Hajal, and M.~Cernak, ``{byol-S: Learning Self-supervised Speech Representations by Bootstrapping},'' in \emph{HEAR: Holistic Evaluation of Audio Representations}.\hskip 1em plus 0.5em minus 0.4em\relax PMLR, 2022, pp. 25--47.

\bibitem{liu2024dinosr}
A.~H. Liu, H.-J. Chang, M.~Auli, W.-N. Hsu, and J.~Glass, ``{DinoSR: Self-distillation and Online Clustering for Self-supervised Speech Representation learning},'' \emph{Advances in Neural Information Processing Systems}, vol.~36, 2024.

\bibitem{10095373}
Q.-S. Zhu, L.~Zhou, J.~Zhang, S.-J. Liu, Y.-C. Hu, and L.-R. Dai, ``{Robust Data2VEC: Noise-Robust Speech Representation Learning for ASR by Combining Regression and Improved Contrastive Learning},'' in \emph{Proceeding of ICASSP}, 2023, pp. 1--5.

\bibitem{caron2018deep}
M.~Caron, P.~Bojanowski, A.~Joulin, and M.~Douze, ``{Deep Clustering for Unsupervised Learning of Visual Features},'' in \emph{Proceedings of ECCV}, 2018.

\bibitem{asano_self-labelling_2020}
Y.~M. Asano, C.~Rupprecht, and A.~Vedaldi, ``Self-{{Labelling Via Simultaneous Clustering}} and {{Representation Learning}},'' in \emph{{{ICLR}}}, 2020.

\bibitem{li_prototypical_2021}
J.~Li, P.~Zhou, C.~Xiong, and S.~C.~H. Hoi, ``Prototypical {{Contrastive Learning}} of {{Unsupervised Representations}},'' in \emph{{{ICLR}}}, 2021.

\bibitem{zhan_online_2020}
X.~Zhan, J.~Xie, Z.~Liu, Y.-S. Ong, and C.~C. Loy, ``Online {{Deep Clustering}} for {{Unsupervised Representation Learning}},'' in \emph{Proceedings of CVPR}, 2020, pp. 6687--6696.

\bibitem{caron_unsupervised_2020}
M.~Caron, I.~Misra, J.~Mairal, P.~Goyal, P.~Bojanowski, and A.~Joulin, ``Unsupervised {{Learning}} of {{Visual Features}} by {{Contrasting Cluster Assignments}},'' \emph{NeurIPS}, vol.~33, pp. 9912--9924, 2020.

\bibitem{chang23_interspeech}
H.-J. Chang, A.~H. Liu, and J.~Glass, ``{Self-supervised Fine-tuning for Improved Content Representations by Speaker-invariant Clustering},'' in \emph{Proceeding of Interspeech}, 2023, pp. 2983--2987.

\bibitem{zhu2022introduction}
X.~Zhu and A.~B. Goldberg, \emph{{Introduction to Semi-Supervised Learning}}.\hskip 1em plus 0.5em minus 0.4em\relax Springer Nature, 2022.

\bibitem{9941371}
X.~Yang, Z.~Song, I.~King, and Z.~Xu, ``{A Survey on Deep Semi-Supervised Learning},'' \emph{IEEE Transactions on Knowledge and Data Engineering}, vol.~35, no.~9, pp. 8934--8954, 2023.

\bibitem{amini2022self}
M.-R. Amini, V.~Feofanov, L.~Pauletto, E.~Devijver, and Y.~Maximov, ``{Self-training: A survey},'' \emph{arXiv:2202.12040}, 2022.

\bibitem{ke2019dual}
Z.~Ke, D.~Wang, Q.~Yan, J.~Ren, and R.~W. Lau, ``{Dual student: Breaking the Limits of the Teacher in Semi-supervised Learning},'' in \emph{Proceedings of CVPR}, 2019, pp. 6728--6736.

\bibitem{sohn2020fixmatch}
K.~Sohn, D.~Berthelot, N.~Carlini, Z.~Zhang, H.~Zhang, C.~A. Raffel, E.~D. Cubuk, A.~Kurakin, and C.-L. Li, ``{Fixmatch: Simplifying Semi-supervised Learning with Consistency and Confidence},'' \emph{NeurIPS}, vol.~33, pp. 596--608, 2020.

\bibitem{chen2021semi}
X.~Chen, Y.~Yuan, G.~Zeng, and J.~Wang, ``{Semi-supervised Semantic Segmentation with Cross Pseudo Supervision},'' in \emph{Proceedings of CVPR}, 2021, pp. 2613--2622.

\bibitem{xie_self-training_2020}
Q.~Xie, M.-T. Luong, E.~Hovy, and Q.~V. Le, ``{Self-training with Noisy Student Improves ImageNet Classification},'' in \emph{Proceedings of CVPR}, 2020, pp. 10\,687--10\,698.

\bibitem{busto2018open}
P.~P. Busto, A.~Iqbal, and J.~Gall, ``{Open Set Domain Adaptation for Image and Action Recognition},'' \emph{IEEE Transactions on Pattern Analysis and Machine Intelligence}, vol.~42, no.~2, pp. 413--429, 2018.

\bibitem{french2018self}
G.~French, M.~Mackiewicz, and M.~Fisher, ``{Self-ensembling for Visual Domain Adaptation},'' in \emph{ICLR}, 2018.

\bibitem{zou2019confidence}
Y.~Zou, Z.~Yu, X.~Liu, B.~Kumar, and J.~Wang, ``{Confidence Regularized Self-training},'' in \emph{Proceedings of CVPR}, 2019, pp. 5982--5991.

\bibitem{zou2018unsupervised}
Y.~Zou, Z.~Yu, B.~Kumar, and J.~Wang, ``{Unsupervised Domain Adaptation for Semantic Segmentation via Class-balanced Self-training},'' in \emph{Proceedings of ECCV}, 2018, pp. 289--305.

\bibitem{tarvainen2017mean}
A.~Tarvainen and H.~Valpola, ``{Mean Teachers are Better Role Models: Weight-Averaged Consistency Targets Improve Semi-supervised Deep Learning Results},'' \emph{NeurIPS}, vol.~30, 2017.

\bibitem{laine2016temporal}
S.~Laine and T.~Aila, ``{Temporal Ensembling for Semi-Supervised Learning},'' in \emph{ICLR}, 2016.

\bibitem{qiao2018deep}
S.~Qiao, W.~Shen, Z.~Zhang, B.~Wang, and A.~Yuille, ``{Deep Co-training for Semi-supervised Image Recognition},'' in \emph{Proceedings of ECCV}, 2018, pp. 135--152.

\bibitem{dong2018tri}
W.~Dong-DongChen and Z.~WeiGao, ``{Tri-net for Semi-supervised Deep Learning},'' in \emph{Proceeding of IJCAI}, 2018, pp. 2014--2020.

\bibitem{reynolds2000speaker}
D.~A. Reynolds, T.~F. Quatieri, and R.~B. Dunn, ``{Speaker Verification using Adapted Gaussian Mixture Models},'' \emph{Digital signal processing}, vol.~10, no. 1-3, pp. 19--41, 2000.

\bibitem{gauvain1994maximum}
J.-L. Gauvain and C.-H. Lee, ``{Maximum a Posteriori Estimation for Multivariate Gaussian Mixture Observations of Markov Chains},'' \emph{IEEE transactions on speech and audio processing}, vol.~2, no.~2, pp. 291--298, 1994.

\bibitem{kenny2007joint}
P.~Kenny, G.~Boulianne, P.~Ouellet, and P.~Dumouchel, ``{Joint Factor Analysis versus Eigenchannels in Speaker Recognition},'' \emph{IEEE Transactions on Audio, Speech, and Language Processing}, vol.~15, no.~4, pp. 1435--1447, 2007.

\bibitem{dehak_front-end_2011}
N.~Dehak, P.~J. Kenny, R.~Dehak, P.~Dumouchel, and P.~Ouellet, ``Front-{End} {Factor} {Analysis} for {Speaker} {Verification},'' \emph{IEEE Transactions on Audio, Speech, and Language Processing}, vol.~19, no.~4, pp. 788--798, 2011.

\bibitem{7404779}
D.~Snyder, D.~Garcia-Romero, and D.~Povey, ``{Time Delay Deep Neural Network-based Universal Background Models for Speaker Recognition},'' in \emph{Proceeding of IEEE Automatic Speech Recognition and Understanding Workshop}, 2015, pp. 92--97.

\bibitem{snyder_x-vectors:_2018}
D.~Snyder, D.~{Garcia-Romero}, G.~Sell, D.~Povey, and S.~Khudanpur, ``X-vectors: {{Robust DNN Embeddings}} for {{Speaker Recognition}},'' in \emph{Proceeding of ICASSP}, 2018, pp. 5329--5333.

\bibitem{cai2018exploring}
W.~Cai, J.~Chen, and M.~Li, ``{Exploring the Encoding Layer and Loss Function in End-to-End Speaker and Language Recognition System},'' in \emph{{Proceeding of The Speaker and Language Recognition Workshop (Odyssey)}}, 2018.

\bibitem{zhou_resnext_2021}
T.~Zhou, Y.~Zhao, and J.~Wu, ``{{ResNeXt}} and {{Res2Net Structures}} for {{Speaker Verification}},'' in \emph{Proceeding of IEEE Spoken Language Technology Workshop}, 2021, pp. 301--307.

\bibitem{desplanques_ecapa-tdnn_2020}
B.~Desplanques, J.~Thienpondt, and K.~Demuynck, ``{{ECAPA-TDNN}}: {{Emphasized Channel Attention}}, {{Propagation}} and {{Aggregation}} in {{TDNN Based Speaker Verification}},'' in \emph{Proceeding of Interspeech}, 2020, pp. 3830--3834.

\bibitem{zhang_mfa-conformer_2022}
Y.~Zhang, Z.~Lv, H.~Wu, S.~Zhang, P.~Hu, Z.~Wu, H.-y. Lee, and H.~Meng, ``{{MFA-Conformer}}: {{Multi-scale Feature Aggregation Conformer}} for {{Automatic Speaker Verification}},'' in \emph{Proceeding of Interspeech}, 2022, pp. 306--310.

\bibitem{10095433}
D.~Liao, T.~Jiang, F.~Wang, L.~Li, and Q.~Hong, ``{Towards A Unified Conformer Structure: from ASR to ASV Task},'' in \emph{Proceeding of ICASSP}, 2023, pp. 1--5.

\bibitem{10572375}
D.~Cai and M.~Li, ``{Leveraging ASR Pretrained Conformers for Speaker Verification Through Transfer Learning and Knowledge Distillation},'' \emph{IEEE/ACM Transactions on Audio, Speech, and Language Processing}, vol.~32, pp. 3532--3545, 2024.

\bibitem{baevski_wav2vec_2020}
A.~Baevski, H.~Zhou, A.~Mohamed, and M.~Auli, ``wav2vec 2.0: {{A Framework}} for {{Self-Supervised Learning}} of {{Speech Representations}},'' \emph{Advances in Neural iInformation Processing Systems}, vol.~33, pp. 12\,449--12\,460, 2020.

\bibitem{hsu_hubert_2021}
W.-N. Hsu, Y.-H.~H. Tsai, B.~Bolte, R.~Salakhutdinov, and A.~Mohamed, ``Hubert: {{How Much Can}} a {{Bad Teacher Benefit ASR Pre-Training}}?'' in \emph{{{Proceeding of ICASSP}}}, 2021, pp. 6533--6537.

\bibitem{9814838}
S.~Chen, C.~Wang, Z.~Chen, Y.~Wu, S.~Liu, Z.~Chen, J.~Li, N.~Kanda, T.~Yoshioka, X.~Xiao, J.~Wu, L.~Zhou, S.~Ren, Y.~Qian, Y.~Qian, J.~Wu, M.~Zeng, X.~Yu, and F.~Wei, ``{WavLM: Large-Scale Self-Supervised Pre-Training for Full Stack Speech Processing},'' \emph{IEEE Journal of Selected Topics in Signal Processing}, vol.~16, no.~6, pp. 1505--1518, 2022.

\bibitem{fan_exploring_2021}
Z.~Fan, M.~Li, S.~Zhou, and B.~Xu, ``{Exploring wav2vec 2.0 on Speaker Verification and Language Identification},'' in \emph{Proceeding of Interspeech}, 2021, pp. 1509--1513.

\bibitem{vaessen_fine-tuning_2022}
N.~Vaessen and D.~A. {van Leeuwen}, ``{Fine-Tuning wav2vec2 for Speaker Recognition},'' in \emph{{{Proceeding of ICASSP}}}, 2022, pp. 7967--7971.

\bibitem{9714360}
J.~Li, K.~Zheng, J.~Yao, L.~Gao, and D.~Hong, ``{Deep Unsupervised Blind Hyperspectral and Multispectral Data Fusion},'' \emph{IEEE Geoscience and Remote Sensing Letters}, vol.~19, pp. 1--5, 2022.

\bibitem{10233913}
J.~Li, K.~Zheng, W.~Liu, Z.~Li, H.~Yu, and L.~Ni, ``{Model-Guided Coarse-to-Fine Fusion Network for Unsupervised Hyperspectral Image Super-Resolution},'' \emph{IEEE Geoscience and Remote Sensing Letters}, vol.~20, pp. 1--5, 2023.

\bibitem{10504844}
J.~Li, K.~Zheng, L.~Gao, L.~Ni, M.~Huang, and J.~Chanussot, ``{Model-Informed Multistage Unsupervised Network for Hyperspectral Image Super-Resolution},'' \emph{IEEE Transactions on Geoscience and Remote Sensing}, vol.~62, pp. 1--17, 2024.

\bibitem{thienpondtidlab}
J.~Thienpondt, B.~Desplanques, and K.~Demuynck, ``{The IDLAB VoxCeleb Speaker Recognition Challenge 2020 System Description},'' in \emph{{{VoxSRC workshop}}}, 2020.

\bibitem{10314722}
B.~Han, Z.~Chen, and Y.~Qian, ``{Self-Supervised Learning With Cluster-Aware-DINO for High-Performance Robust Speaker Verification},'' \emph{IEEE/ACM Transactions on Audio, Speech, and Language Processing}, vol.~32, pp. 529--541, 2024.

\bibitem{zhao2024prototype}
Z.~Zhao, Z.~Li, X.~Zhang, W.~Wang, and P.~Zhang, ``{Prototype Division for Self-Supervised Speaker Verification},'' \emph{IEEE Signal Processing Letters}, vol.~31, pp. 880--884, 2024.

\bibitem{tao2023self}
R.~Tao, K.~A. Lee, R.~K. Das, V.~Hautam{\"a}ki, and H.~Li, ``{Self-Supervised Training of Speaker Encoder with Multi-Modal Diverse Positive Pairs},'' \emph{IEEE/ACM Transactions on Audio, Speech, and Language Processing}, vol.~31, pp. 1706--1719, 2023.

\bibitem{tao_self-supervised_2022}
R.~\vspace{0mm}Tao, K.~A. Lee, R.~K. Das, V.~Hautam{\"a}ki, and H.~Li, ``Self-{{Supervised Speaker Recognition}} with {{Loss-Gated Learning}},'' in \emph{{{Proceeding of ICASSP}}}, 2022, pp. 6142--6146.

\bibitem{han_self-supervised_2022}
B.~Han, Z.~Chen, and Y.~Qian, ``Self-{{Supervised Speaker Verification Using Dynamic Loss-Gate}} and {{Label Correction}},'' in \emph{Proceeding of Interspeech}, 2022, pp. 4780--4784.

\bibitem{chen_self-supervised_2023}
H.~Chen, H.~Zhang, L.~Wang, K.~A. Lee, M.~Liu, and J.~Dang, ``Self-{{Supervised Audio-Visual Speaker Representation}} with {{Co-Meta Learning}},'' in \emph{{{Proceeding of ICASSP}}}, 2023, pp. 1--5.

\bibitem{fang2024improving}
Z.~Fang, L.~He, L.~Li, and Y.~Hu, ``{Improving Speaker Verification with Noise-Aware Label Ensembling and Sample Selection: Learning and Correcting Noisy Speaker Labels},'' \emph{IEEE/ACM Transactions on Audio, Speech, and Language Processing}, vol.~32, pp. 2988--3001, 2024.

\bibitem{srivastava_dropout:_2014}
N.~Srivastava, G.~Hinton, A.~Krizhevsky, I.~Sutskever, and R.~Salakhutdinov, ``Dropout: {{A Simple Way}} to {{Prevent Neural Networks}} from {{Overfitting}},'' \emph{Journal of Machine Learning Research}, vol.~15, no.~1, pp. 1929--1958, 2014.

\bibitem{deng_arcface_2019}
J.~Deng, J.~Guo, N.~Xue, and S.~Zafeiriou, ``{{ArcFace}}: {{Additive Angular Margin Loss}} for {{Deep Face Recognition}},'' in \emph{{{Proceedings of CVPR}}}, 2019, pp. 4685--4694.

\bibitem{moco}
K.~He, H.~Fan, Y.~Wu, S.~Xie, and R.~Girshick, ``{Momentum Contrast for Unsupervised Visual Representation Learning},'' in \emph{Proceedings of CVPR}, 2020, pp. 9729--9738.

\bibitem{cuturi_sinkhorn_2013}
M.~Cuturi, ``Sinkhorn {{Distances}}: {{Lightspeed Computation}} of {{Optimal Transport}},'' \emph{NeurIPS}, vol.~26, 2013.

\bibitem{nagrani_voxceleb:_2017}
A.~Nagrani, J.~S. Chung, and A.~Zisserman, ``Voxceleb: {A} {Large}-{Scale} {Speaker} {Identification} {Dataset},'' in \emph{Proceeding of Interspeech}, 2017, pp. 2616--2620.

\bibitem{chung_voxceleb2:_2018}
J.~S. Chung, A.~Nagrani, and A.~Zisserman, ``Voxceleb2: {{Deep Speaker Recognition}},'' in \emph{Proceeding of Interspeech}, 2018, pp. 1086--1090.

\bibitem{cai_within-sample_2020}
D.~Cai, W.~Cai, and M.~Li, ``Within-{Sample} {Variability}-{Invariant} {Loss} for {Robust} {Speaker} {Recognition} {Under} {Noisy} {Environments},'' in \emph{{Proceeding of ICASSP}}, 2020, pp. 6469--6473.

\bibitem{inoue_semi-supervised_2020}
N.~Inoue and K.~Goto, ``Semi-{{Supervised Contrastive Learning}} with {{Generalized Contrastive Loss}} and {{its Application}} to {{Speaker Recognition}},'' in \emph{{{Proceeding of APSIPA ASC}}}, 2020, pp. 1641--1646.

\bibitem{huh_augmentation_2020}
J.~Kang, J.~Huh, H.~S. Heo, and J.~S. Chung, ``{Augmentation Adversarial Training for Self-Supervised Speaker Representation Learning},'' \emph{IEEE Journal of Selected Topics in Signal Processing}, vol.~16, no.~6, pp. 1253--1262, 2022.

\bibitem{chen_simple_2020}
T.~Chen, S.~Kornblith, M.~Norouzi, and G.~Hinton, ``A {{Simple Framework}} for {{Contrastive Learning}} of {{Visual Representations}},'' in \emph{{{Proceedings of the ICML}}}, 2020, pp. 1597--1607.

\bibitem{musan}
D.~Snyder, G.~Chen, and D.~Povey, ``{MUSAN}: {A} {Music}, {Speech}, and {Noise} {Corpus},'' \emph{arXiv:1510.08484}, 2015.

\bibitem{ko2017study}
T.~Ko, V.~Peddinti, D.~Povey, M.~L. Seltzer, and S.~Khudanpur, ``{A Study on Data Augmentation of Reverberant Speech for Robust Speech Recognition},'' in \emph{Proceeding of ICASSP}, 2017, pp. 5220--5224.

\bibitem{he_deep_2016}
K.~He, X.~Zhang, S.~Ren, and J.~Sun, ``Deep {{Residual Learning}} for {{Image Recognition}},'' in \emph{{{Proceedings of CVPR}}}, 2016, pp. 770--778.

\bibitem{nist_nist_2016}
\BIBentryALTinterwordspacing
``{{NIST}} 2016 {{Speaker Recognition Evaluation Plan}},'' 2016. [Online]. Available: \url{https://www.nist.gov/system/files/documents/2016/10/07/sre16_eval_plan_v1.3.pdf}
\BIBentrySTDinterwordspacing

\bibitem{asano_labelling_2020}
Y.~M. Asano, M.~Patrick, C.~Rupprecht, and A.~Vedaldi, ``Labelling {{Unlabelled Videos}} from {{Scratch}} with {{Multi}}-{{Modal Self}}-{{Supervision}},'' \emph{NeurIPS}, vol.~33, pp. 4660--4671, 2020.

\bibitem{munkres1957algorithms}
J.~Munkres, ``{Algorithms for the Assignment and Transportation Problems},'' \emph{Journal of the society for industrial and applied mathematics}, vol.~5, no.~1, pp. 32--38, 1957.

\bibitem{mun2021snu}
S.~H. Mun, M.~H. Han, and N.~S. Kim, ``{SNU-HIL System for the VoxCeleb Speaker Recognition Challenge 2021},'' in \emph{{{VoxSRC workshop}}}, 2021.

\bibitem{chen_unsupervised_2023}
Z.~Chen, J.~Wang, W.~Hu, L.~Li, and Q.~Hong, ``Unsupervised {{Speaker Verification Using Pre-Trained Model}} and {{Label Correction}},'' in \emph{{{Proceeding of ICASSP}}}, 2023, pp. 1--5.

\end{thebibliography}
\end{document}